\pdfoutput=1

\documentclass[11pt,twoside,a4paper,cmspaper,final,collab]{cms-tdr}
\def\svnVersion{59192:59198}\def\cmsCernNo{2011-060}\def\cmsCernDate{\today}\def\cmsMessage{Submitted to the European Physical Journal C}

\begin{document}\cmsNoteHeader{TOP-10-002}

\hyphenation{had-ron-i-za-tion}
\hyphenation{cal-or-i-me-ter}
\hyphenation{de-vices}
\RCS$Revision: 59198 $
\RCS$HeadURL: svn+ssh://alverson@svn.cern.ch/reps/tdr2/papers/TOP-10-002/trunk/TOP-10-002.tex $
\RCS$Id: TOP-10-002.tex 59198 2011-06-05 14:05:22Z alverson $
\cmsNoteHeader{TOP-10-002} 
\title{Measurement of the ${\rm t{\bar t}}$ Production Cross
Section\\ in pp Collisions at $\sqrt{s}=7\,\rm{TeV}$ \\
using the Kinematic Properties \\of
Events with Leptons and Jets}

\address[neu]{Northeastern University}
\address[fnal]{Fermilab}
\address[cern]{CERN}
\author[neu]{George Alverson}\author[fnal]{Lucas Taylor}\author[cern]{A. Cern Person}

\date{\today}

\abstract{
A measurement of the ${\rm t{\bar t}}$ production cross section
in proton-proton
collisions at a centre-of-mass energy of 7~TeV
has been performed at the LHC with the CMS detector.
The analysis uses a data sample corresponding to an integrated
luminosity of 36~pb$^{-1}$
and
is based on the reconstruction of the
final state with
one isolated, high transverse-momentum electron or muon
and three or more hadronic jets. The kinematic properties
of the events are used to separate the
${\rm t{\bar t}}$ signal from W$+$jets and QCD multijet
background events.
The measured cross section is
$173^{+39}_{-32}$~(stat. + syst.)~pb,
consistent with standard model expectations.
}

\hypersetup{%
pdfauthor={CMS Collaboration},%
pdftitle={Measurement of the Top antitop Production Cross Section in pp Collisions at sqrt(s)=7 TeV using the Kinematic Properties of Events with Leptons and Jets},%
pdfsubject={CMS},%
pdfkeywords={CMS, LHC, physics, top quark, cross section}}

\maketitle 

\newcommand{\met}{\mbox{$\rlap{\kern0.15em/}E_\mathrm{T}$}}

\section{Introduction}
\label{sec1:intro}
The top quark occupies a unique position within the standard model.
With a mass roughly that of a tungsten atom, it is the 
only quark heavy enough to decay before forming bound
states with other quarks. 
Its large mass has inspired
numerous theoretical models in which the top quark 
plays a special role in the generation of mass or in the physics of new,
undiscovered particles.
The top quark often acts as either a direct contributor to new physics
or an important background in new-particle searches in these models.  

In hadron colliders, 
top-quark production is dominated by the production
of ${\rm t{\bar t}}$ pairs~\cite{top-prospects}.
At the Tevatron,
where the top quark was discovered 
in 1995~\cite{top-discovery-cdf,top-discovery-d0},
${\rm t{\bar t}}$
pairs are predominantly produced through quark-antiquark annihilation.
In contrast,
the ${\rm t{\bar t}}$ production mechanism at the 
Large Hadron Collider (LHC)~\cite{lhc} is expected to be
dominated by the gluon fusion process.
Measurements of the ${\rm t{\bar t}}$ cross section
at the LHC can provide
important tests of our understanding of the top-quark
production mechanism and can also be used in searches for new physics.

In the standard model, a top quark decays nearly 100\%
of the time to a W boson and a b~quark.
The decay of a 
${\rm t{\bar t}}$ pair is categorized by the 
decay of the W bosons produced by
the pair. Thus the channel 
in which both W bosons
decay to leptons is referred to as 
the ``dilepton'' channel, 
and the channel in which one W decays 
to leptons and the other to quark jets is the ``lepton$+$jets'' channel.
The channel in which both W bosons decay to jets is
called the ``all hadronic'' channel.  
A further categorization of the decays is made by specifying
the flavour of the charged lepton(s) 
produced from the W decays. 
For the purposes of this paper, the "lepton$+$jets" channel refers
only to decays in which the charged lepton is either an electron or a
muon.

\sloppy{
The next-to-leading-order (NLO) corrections to the top-quark pair
production cross section at
hadron colliders have been calculated both with
the full top-quark spin dependence~\cite{Bernreuther1,Bernreuther2}
and without this dependence~\cite{Nason,Beenakker}. 
A complete analytic result for the NLO partonic cross section has
only been published recently~\cite{Czakon}.
Approximations of a full next-to-next-to-leading-order (NNLO)
calculation have also
been obtained by various groups \cite[and references
  therein]{Aliev,Langenfeld,Kidonakis, Cacciari}. The ${\rm t{\bar
t}}$ production cross section at the LHC has been previously 
measured in the dilepton channel by the
Compact Muon Solenoid (CMS) experiment~\cite{CMS_dilepton}
and in the combined
dilepton and lepton$+$jets channels by the ATLAS
experiment~\cite{ATLAS_top_sigma}. 
These measurements agree with recent NLO and
with approximate NNLO calculations.
}

\sloppy{
In this paper we present a measurement of the 
cross section for ${\rm t\bar{t}}$ production in proton-proton collisions 
at ${\sqrt{s}=7\,\rm{TeV}}$ with the CMS
detector, using the electron$+$jets and muon$+$jets final
states. 
Although there are two jets from hadronization of the b~quarks in
these final states, in this analysis no requirement is made
on the presence of b~jets. Instead, the kinematic properties
of the events are used to select the ${\rm t{\bar t}}$
signal. It is important to measure the ${\rm t{\bar t}}$
cross section both with and without a requirement on the presence of 
b~jets, as new physics could contribute differently 
in each case.}

A brief overview of the CMS detector is provided in Section~\ref{sec:CMS}
of this paper, followed by a discussion of event reconstruction
procedures in Section~\ref{sec:reco}.  The
selection criteria applied to the data are described in
Section~\ref{sec:select} and the processes used to simulate signal and
background events are described in Section~\ref{sec:sim}.
Section~\ref{sec:xsec} details the method used
to extract a measurement of the ${\rm t{\bar t}}$ cross section from the
selected events, as well as the calculation of statistical and
systematic uncertainties on the result.  Section~\ref{sec:conclusion}
summarizes the result and compares the measurement with recent
perturbative QCD calculations.

\section{CMS Detector}
\label{sec:CMS}
The central feature of the CMS apparatus is a 3.8~T magnetic field
produced by a superconducting solenoid of 6~m internal diameter.
Within the field volume are the silicon pixel and strip
trackers, the crystal electromagnetic calorimeter (ECAL), and the
hadron calorimeter (HCAL). 
The inner tracker 
measures charged particles 
within the range~$|\eta| < 2.5$, where~$\eta$ indicates detector
pseudorapidity. It consists of 1440 silicon pixel and 15\,148
silicon strip detector modules and is located within the axial
magnetic field. It provides an impact parameter resolution
of~$\sim$~15~$\mu$m and a transverse 
momentum~($p_{\rm T}$) resolution of $\sim$~1.5\% for 100~GeV particles.
The ECAL consists of nearly 76\,000 lead tungstate crystals that
provide coverage in pseudorapidity of~$\vert \eta \vert< 1.48$ in the
ECAL barrel region and~$1.48 <\vert \eta \vert < 3.0$ in the two endcaps. A
preshower detector consisting of two planes of silicon
sensors interleaved with a total of 3~radiation lengths of lead is located in
front of the endcaps. The ECAL energy resolution is 3\% or better for
the range of
electron energies relevant for this analysis. 

The HCAL is composed of layers of plastic
scintillator within a brass/stainless steel absorber, covering the
region $|\eta|<3.0$. A calorimeter composed of quartz fibres embedded
in a steel absorber extends the forward HCAL coverage beyond the
solenoid volume, to $|\eta|<5.0$.
In the region~$\vert \eta \vert< 1.74$, the HCAL cells have widths of 0.087 in
both pseudorapidity and azimuth ($\phi$). In
the~$(\eta,\phi)$ plane, for~$\vert \eta \vert< 1.48$, the HCAL cells
match the underlying~$5 \times 5$ ECAL crystal arrays to form
calorimeter towers
projecting radially outwards from close to the nominal interaction
point. At larger values of~$\vert \eta \vert$, the size of the towers
increases, and the matching ECAL arrays contain fewer crystals. 

Muons are measured in the pseudorapidity window~$|\eta|< 2.4$,
with detection planes made of drift tubes, cathode
strip chambers, and resistive plate chambers. Matching the muons to
the tracks measured in the silicon tracker results in a
transverse-momentum resolution between 1\% and 5\% for~$p_{\rm T}$
values up to 1~\rm{TeV}.
A two-tier trigger system selects the
most interesting pp collision events for use in physics analyses. A
more detailed description of the CMS detector can be found
elsewhere~\cite{cms}.

\section{Event Reconstruction}
\label{sec:reco}

The reconstruction of 
electrons~\cite{Electronreco, ElectronPAS} uses information
from the pixel detector, the silicon strip tracker, and the
electromagnetic calorimeter. 
The amount of
material before the ECAL 
in the CMS detector ranges from 0.4 to 1.6 radiation lengths,
depending on~$\eta$, and an electron may lose a considerable fraction of
its energy through bremsstrahlung in passing through this material. 
The energy deposited in the calorimeter may be spread over a wide
range in $\phi$ compared to the initial direction of the electron.
To account for this in the 
reconstruction of electron candidates, 
clusters of calorimeter energy 
deposition (``superclusters'') from a narrow fixed range in $\eta$ and a
variable range in $\phi$ are formed.
Starting from these superclusters, corresponding hits
in at least two layers of the pixel tracker 
capable of acting as seeds for electron trajectory candidates are required. 
Energy loss through bremsstrahlung
leads to non-Gaussian contributions 
to fluctuations in the calorimeter and tracking measurements. 
Therefore, the seeding and 
building of tracks is done using
dedicated algorithms designed to handle these fluctuations.
The final fit of the trajectories relies on a Gaussian sum filter
that is a non-linear
generalization of the Kalman filter with weighted sums of Gaussians instead of 
a single Gaussian.

Multiple reconstruction algorithms exist to identify muon candidates
in CMS from hits in the silicon tracking system and signals in the
muon system~\cite{MUOPAS,cosmic_muon}. 
Since 
muons are typically the only particles reaching the muon chambers,
the ``stand-alone muon'' 
reconstruction algorithm starts from track segments detected in the
innermost layers
of the muon chambers. Additional hits in surrounding layers are
then added. 
Finally, muon tracks are 
propagated back to the interaction point. A second reconstruction 
algorithm (``tracker muons'')
starts from tracks found in the tracking system and then 
associates them with
compatible signals in the calorimeters and the muon chambers.
A third algorithm (``global muons'') starts from stand-alone muons and
then makes a global fit to the consistent hits in the tracker
and the muon system. 
Muons used in this analysis 
are required to be identified by both the tracker muon 
and the global muon algorithms.

Hadronic jets are reconstructed from individual particles 
whose identities and energies have been determined via a
particle-flow technique~\cite{PFT-09-001} that combines information
from all
subdetectors: charged tracks in the tracker and energy deposits in the
electromagnetic and hadronic calorimeters, as well as signals in the
preshower detector and the muon system. The energy calibration is
performed separately for each particle type. 
All particles
found by the particle-flow algorithm are clustered into
particle-flow jets~\cite{PFT-10-002, JME-10-003} by using each particle's
direction at the interaction vertex and the 
anti-$k_{\rm{ T}}$ jet
clustering algorithm~\cite{antikt} with the distance parameter
R~=~0.5, as implemented in \textsc{FastJet} version
2.4~\cite{fastjet}.

Since most jet constituents are identified and
reconstructed with nearly the correct energy by the particle-flow algorithm,
only small residual jet energy corrections must be applied to each jet. 
These corrections are between 5\%\ and 10\%\ of the jet energy and
were obtained as a function of $p_{\rm{T}}$ and
$\eta$ from the \textsc{Geant4}-based CMS Monte Carlo
simulation~(v. 9.2 Rev01)~\cite{Geant4} and early collision data.
The factors also include corrections for  
small discrepancies observed between
the simulation and the data.

The missing transverse energy (\met) is defined as the magnitude of
the negative vector sum of the transverse energies ($E_{\rm{T}}$) of all the
particles found by the particle-flow algorithm. 
A decay of a 
${\rm t{\bar t}}$ pair via the lepton$+$jets channel is expected to exhibit
significant missing tranverse energy because of the undetected neutrino
from the leptonically decaying~W. Distributions of this
variable are used in likelihood fits to measure the ${\rm t{\bar t}}$
signal and to distinguish it from various backgrounds, as discussed in
the following sections. 

\section{Data Set and Event Selection}
\label{sec:select}

The data discussed in this paper were collected in the 
period April to November 2010 from proton-proton collisions
at $\sqrt{s} = 7$~TeV and 
correspond to an integrated luminosity of  
$36 \pm 1\,\rm{pb}^{-1}$\cite{EWK10004,LUMI_Update}. The trigger providing 
the data sample used in this analysis
is based on the presence of at least one charged lepton, either an
electron or a muon. Because the peak instantaneous luminosity increased
throughout the data-taking period, the minimum transverse momentum
$p_{\mathrm{T}}$ of the muon
required in the trigger ranged from 9 to 15~$\mathrm{GeV}$, and the
minimum $E_{\rm{T}}$ required in the
trigger for electrons similarly ranged
from 10 to 22~$\mathrm{GeV}$.
This data sample is
used for the selection of the signal region and the selection of
signal-depleted control regions used for studies related to background
processes.  Trigger efficiencies are determined from the data
using $Z$-boson events and then corrected for the differences in the
efficiencies between the $Z$-boson events and ${\rm t{\bar t}}$ events,
using the simulated samples described below. 
Events are required to have at least one primary
pp~interaction vertex, where vertices are identified by applying an
adaptive fit to clusters of reconstructed tracks~\cite{trkpas}.
The primary vertex must be within~$\pm 24$~cm 
of the nominal interaction point in the direction along the proton
beams.  The distance between the primary vertex and the nominal
interaction point in the plane perpendicular to the beam direction
must be less than $2$~cm.

In the event selection for the electron$+$jets channel, 
at least one electron with transverse energy greater than
30~$\mathrm{GeV}$ and $|\eta |$ less than~2.5 is required.
Electrons from the transition region between the
ECAL barrel and endcap, $1.44 < |\eta_{sc}| < 1.57$,
are excluded, where $\eta_{sc}$ is the pseudorapidity of the ECAL supercluster.
The energy of the HCAL cell mapped to the supercluster must be less
than 2.5\% of the total calorimeter energy associated with the
supercluster.  Additional requirements are made on the shower
shape and the angular separation between the ECAL supercluster and the
matching track.  
Electron tracks must extrapolate to within~0.02~cm of the interaction vertex
in the plane perpendicular to the proton beams and to within~1~cm in the
direction along the beams. Electron candidates that lack hits in the
inner layers of the tracking system are assumed to be the product of
photon conversions and are discarded. 

Since the electron from the W boson in a top-quark decay is expected
to be isolated from other high-$p_{\rm{T}}$ particles in the event,
electrons are required to have a relative isolation ($I_{\rm{rel}}$) 
smaller than 0.1, where relative isolation is defined as the scalar
sum of the transverse momenta of tracks with $p_{\rm{T}}>1$~GeV and
all calorimeter energy in a cone of $\Delta R \equiv
\sqrt{(\Delta\phi)^2 + (\Delta\eta)^2} < 0.3$ around the electron,
divided by the electron $p_{\rm{T}}$. Here $\Delta\phi$ ($\Delta\eta$)
is the difference in azimuthal angle (pseudorapidity) between the
electron and the track or calorimeter cell. Contributions to the sum
due to the electron itself are removed. Events containing multiple
electron candidates are rejected if any combined
dielectron invariant mass lies within 15~$\mathrm{GeV}$ of the
$Z$-boson mass.

In the event selection for the muon$+$jets channel,
muons are 
required to have at least a minimum number of hits
in both the silicon tracking system and the muon chambers.
The muon must have transverse momentum greater than 20~GeV
and must lie within the muon trigger acceptance ($|\eta |<2.1$). The
muon must be separated from any 
selected jet (defined below)
in the event by $\Delta R > 0.3$.
Muons must extrapolate to within 0.02~cm of the interaction vertex
in the plane perpendicular to the beams and to within 1~cm in the
direction along the beams. The muon from the W~boson in a top decay is
also expected to be isolated from other high-$p_{\rm{T}}$ 
particles in the
event, and thus muons are required to have relative isolation
smaller than 0.05, where the muon relative isolation is defined
analogously to the
electron $I_{\rm{rel}}$.
Exactly one muon passing all these criteria is required. Events
containing a separate, more loosely-defined muon,
as well as the highly-energetic isolated muon described above, are rejected.

$Z$-boson events are used to study lepton trigger,
identification, and isolation efficiencies in the data. A high purity
$Z$-boson sample is extracted from data by requiring two oppositely-charged like-flavour
leptons with a combined invariant mass within 
15~$\mathrm{GeV}$ of the $Z$-boson
mass. The two-electron final state
suffers from background contamination due to hadronic jets
misidentified as electrons. This contamination is modeled using
events containing a pair of like-charged electrons whose invariant
mass falls within the $Z$-boson mass window. The identification efficiency for
isolated electrons is determined to be $0.75$, with combined
statistical and systematic uncertainty (stat.+syst.) of  $\pm0.01$.
The trigger efficiency for such electrons is measured to be $0.982 \pm
0.001\,\rm{(stat.)}$.  Using 
$Z \rightarrow \mu\mu$ events, the efficiency of finding an isolated
muon is measured to be $0.880 \pm 0.002\,\rm{(stat.+syst.)}$, and the efficiency for
triggering on muons passing all selection cuts is found to be $0.922
\pm 0.002\,\rm{(stat.)}$. 

Selected jets are required to have a jet-energy-scale-corrected $p_{\rm{T}} >
30$~$\mathrm{GeV}$, $|\eta |<$~2.4, and must be separated by
$\Delta\mathrm{R}>0.3$ from isolated electrons and
$\Delta\mathrm{R}>0.1$ from isolated muons in order to remove jet
candidates produced by charged leptons. Both the muon$+$jets and
electron$+$jets analyses require
that events contain at least three jets.  Selected events are
grouped into subsamples based on their jet multiplicity, so that events
containing exactly three jets are separated from those containing
four or more jets. 
A requirement on \met~is not included in the event selection,
as fits to the distribution of this observable are used to separate
the ${\rm t{\bar t}}$ signal from the backgrounds.

Dilepton ${\rm t{\bar t}}$ decays are removed by discarding events
that contain both a high-$p_{\rm{T}}$ electron and a high-$p_{\rm{T}}$ muon.
In the electron$+$jets event selection, events containing any muon
with $p_{\rm{T}}>10$~GeV, $|\eta|<2.5$, and relative isolation~$<
0.2$ are rejected. In the muon$+$jets 
event selection, events must contain no electron candidates 
with $E_{\rm{T}}>15$~$\mathrm{GeV}$ and $I_{\rm{rel}}<0.2$.

\section{Simulation of Signal and Background Events}
\label{sec:sim}

We determine the efficiency for selecting lepton$+$jets signal events
by using a simulated ${\rm t{\bar t}}$
event sample. We perform the simulation of signal ${\rm t{\bar t}}$ events
using \textsc{MadGraph}~(v. 4.4.12)~\cite{Alwall:2007st},
where the top-antitop pairs are generated accompanied by up to three
additional hard jets in the matrix-element calculation.
The factorization and renormalization scales~$Q^2$
for ${\rm t{\bar t}}$ are both set to

\begin{equation}
Q^2=m_{\mathrm{top}}^2+\Sigma p_{\rm T}^2,
\end{equation}
where $m_{\mathrm{top}}$ is the top-quark mass and $\Sigma p_{\rm T}^2$ is the
sum of the squared transverse momenta of all accompanying hard jets in the
event. \textsc{MadGraph} is also used to generate
background events from electroweak production of single top quarks,
the production of leptonically-decaying W and Z bosons in association
with up to four extra jets, and photon$+$jets processes.
The factorization and
renormalization scales in these events are set in the same manner as for
${\rm t{\bar t}}$ events, with the appropriate boson mass replacing
the top-quark mass. The parton-level results generated by
\textsc{MadGraph} are next processed with
\textsc{Pythia}~(v. 6.420)~\cite{Sjostrand:2006za} to provide showering of the
generated particles. Shower matching is done by applying the MLM
prescription~\cite{MLM_match}. 
Events are then passed through the
\textsc{Geant4}-based CMS detector simulation.

In addition to the simulated events generated with \textsc{MadGraph},
several QCD multijet
samples have been produced using \textsc{Pythia}. Samples enriched in
electrons (muons) are used to provide a preliminary estimate on the QCD
background in the electron$+$jets (muon$+$jets) channel. The final QCD
background contribution is taken directly from data, as is explained
in Section~\ref{sec:xsec}.

The ${\rm t{\bar t}}$ NLO production cross section has been
calculated as 
$\sigma_{\rm t{\bar t}}=157^{+23}_{-24} \rm\ pb$, using
MCFM~\cite{mcfm,mcfm:tt}.
For
both ${\rm t{\bar t}}$ and single-top-quark
production (described below), renormalization and factorization scales were set to
$Q^2=(172.5$~GeV$)^2$.
The uncertainty in the cross section due to uncertainties in these scales
is determined by
varying the scales by factors of 4 and 0.25 around
their nominal values.
Contributions to the cross section uncertainty from the parton
distribution functions (PDF) and the value of $\alpha_S$
are determined following the
procedures from the MSTW2008~\cite{mstw08}, CTEQ6.6~\cite{cteq_2010},
and
NNPDF2.0~\cite{nnpdf} sets. The uncertainties are then combined according to
the PDF4LHC prescriptions~\cite{pdf4lhc}.

The t-channel single-top-quark NLO cross section (multiplied by
the leptonic branching
fraction of the W~boson)
has been
determined as $\sigma_{\rm t} = 21.0^{+1.1}_{-1.0} \rm\ pb$ using
MCFM~\cite{mcfm,mcfm:t:1,mcfm:t:2,mcfm:t:3}, where the uncertainty is defined
similarly
as for ${\rm t{\bar t}}$ production.
The inclusive single-top-quark associated production (tW) NLO
cross section of $\sigma_{\rm{tW}}=10.6\pm0.8 \rm\ pb$~\cite{mcfm:t:2}
has been used. Both cross sections
include the production of single top and single antitop
quarks. The s-channel single-top production cross section is small compared to
the t-channel and tW~production cross sections and is treated as negligible in this analysis.

The NNLO production cross section for W bosons decaying into
leptons has been determined to be $\sigma_{{\rm W}\rightarrow \ell\nu} = 31.3 \pm
1.6 \rm\ nb$ using FEWZ~\cite{fewz}.
Its uncertainty was
determined in a similar manner as for top-quark pair production.
Finally, the Drell-Yan dilepton ($\ell\ell$) production cross section
at NNLO has been calculated using FEWZ as $\sigma_{{\rm Z}/\gamma^*\rightarrow
\ell\ell}({\rm m}_{\ell\ell}>20\,\mathrm{GeV}) = 5.00 \pm 0.27 \rm\ nb$ and
$\sigma_{{\rm Z}/\gamma^*\rightarrow \ell\ell} ({\rm m}_{\ell\ell}>50\,\mathrm{GeV}) =
3.05 \pm 0.13 \rm\ nb$.  Backgrounds due to diboson production have been
ignored given their relatively small expected contribution to the
lepton$+$jets event yield.

\begin{figure}[h!]
 \centering
   \includegraphics[width=0.49\textwidth]{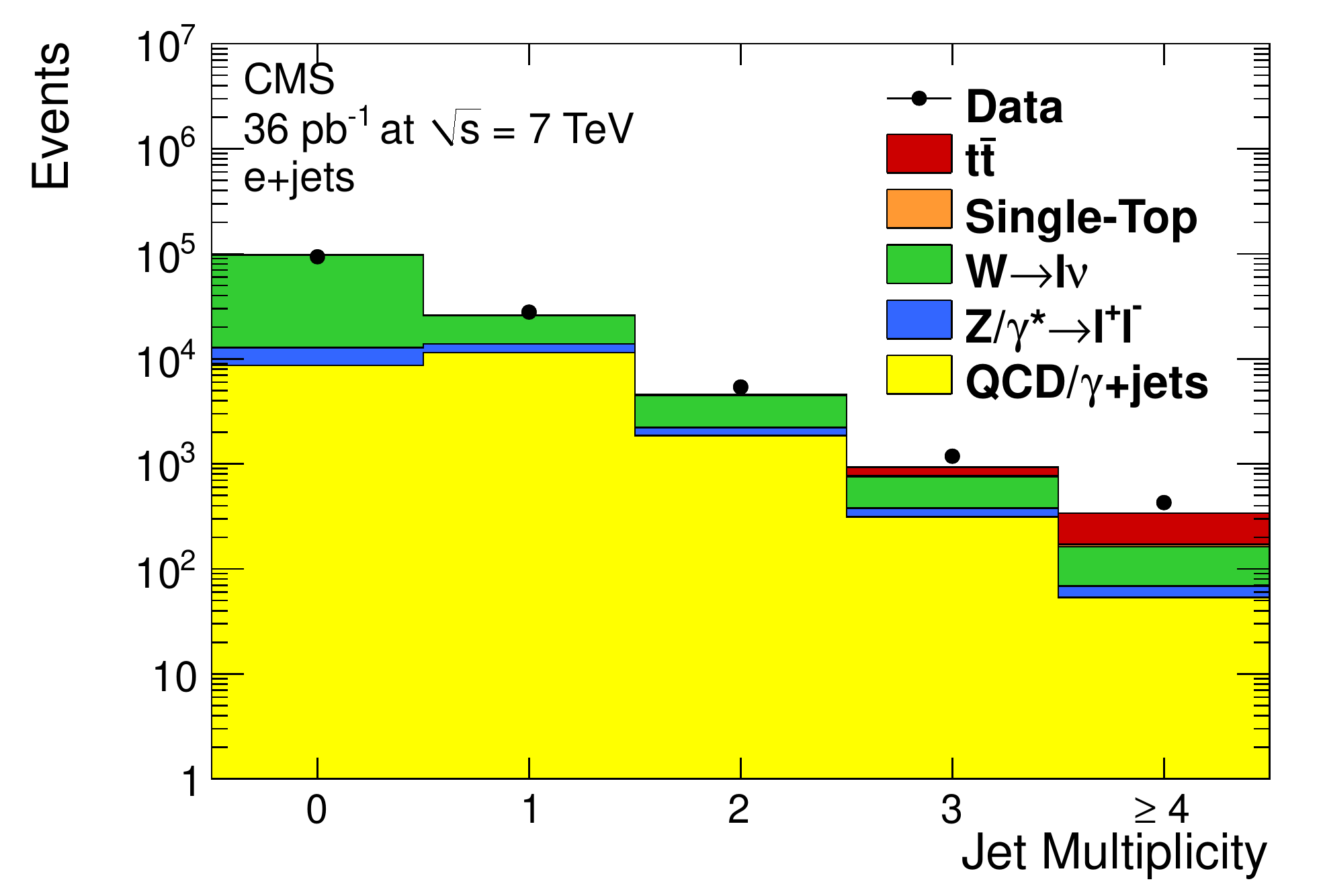}
   \includegraphics[width=0.49\textwidth]{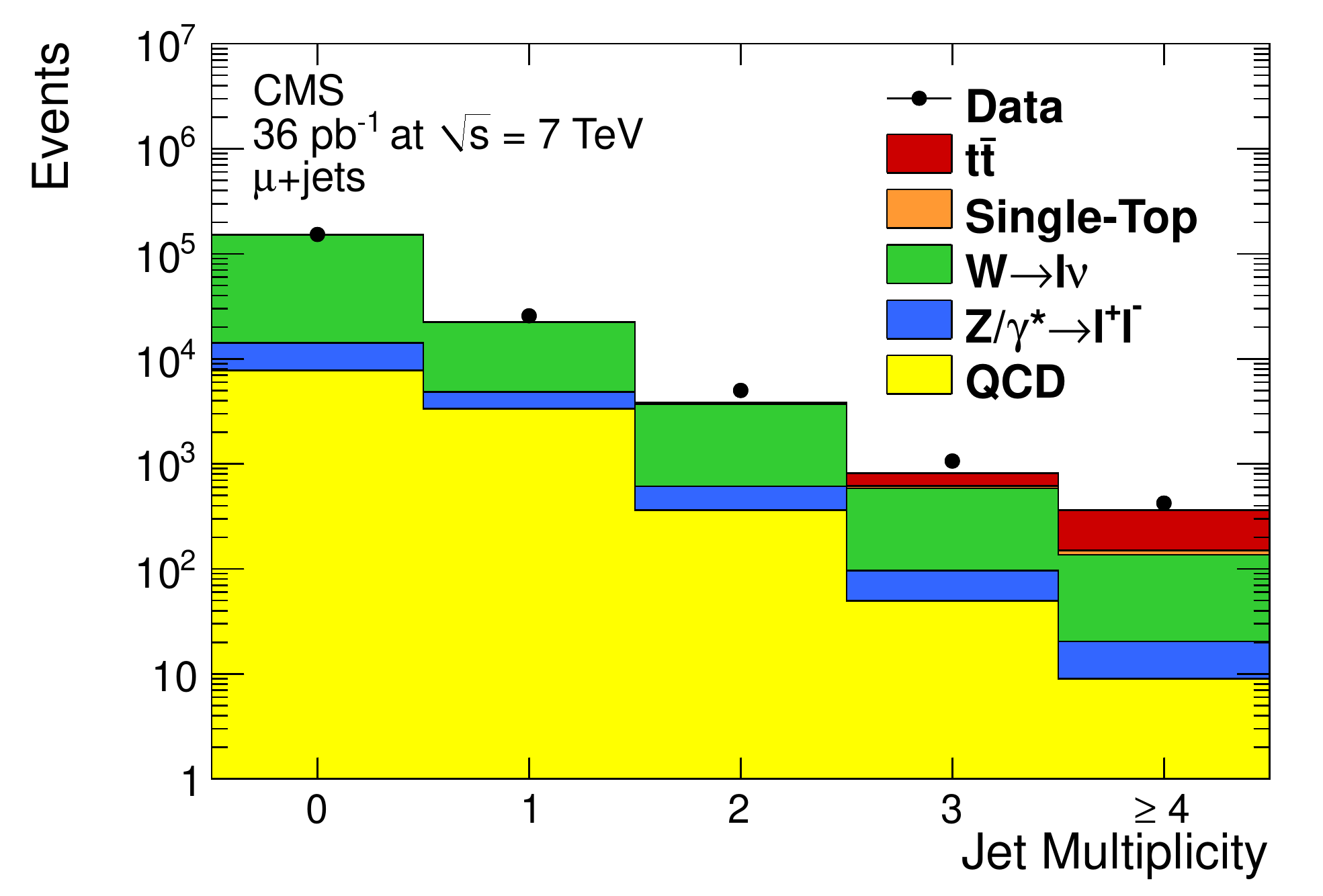}
   \caption{Observed numbers of events from data and simulation as a
     function of jet multiplicity in the (left) electron$+$jets and (right)
muon$+$jets selected samples.  Yields are calculated after applying the respective event
selections and omitting any requirement on the number of jets.  }
 \label{fig:Njet}
\end{figure}

In a simulated sample of ${\rm t{\bar t}}$ events in
which all top-quark decay modes are included, the electron$+$jets
selection efficiency is found to be 5.7\%, while the muon$+$jets
selection efficiency is
7.2\%. The selected simulated signal events in each mode are dominated by ${\rm
t{\bar t}}$ decays to electron$+$jets and muon$+$jets, respectively,
although ${\rm t{\bar t}}$ decays containing tau leptons also
contribute.
Table~\ref{tab:Eventyields} and Fig.~\ref{fig:Njet} give the observed
numbers of events in both channels after applying the event selection
procedures described above to the pp collision data set. The numbers
of events predicted by the simulation for the different physics
processes are also listed. Predicted yields are calculated by multiplying
selection efficiencies for each process, as determined from
simulation, by the appropriate NLO or NNLO cross sections and the
total integrated luminosity of $36$~pb$^{-1}$.
As shown in Fig.~\ref{fig:Njet}, the fractional contribution to the event yield
from ${\rm t{\bar t}}$ signal events is negligible in events with zero
jets, but dominates as jet multiplicity increases.
More events are observed in the data than
predicted by the simulation, indicating that
either the signal cross section or the background cross
sections, or possibly both, are larger than expected.
Our method for determining
the number of signal and background events in
the data is described in the next section.

\begin{table*}[h!]
 \begin{center}
\caption{Predicted event yields for the electron$+$jets
and muon$+$jets event selections. The event yields
from the simulation are normalized to an
integrated luminosity of $36\,\mathrm{pb}^{-1}$. The quoted
uncertainties account for
the limited number of simulated events, the uncertainty on the calculated
cross section (where available),
the uncertainties on the lepton selection and trigger
efficiency correction factors, and a 4\% uncertainty on
the luminosity. The penultimate row lists the
totals from the simulation, and the last row shows the number of
observed events.}
\label{tab:Eventyields}
 \begin{minipage}[c]{170mm}
 \begin{center}
   \begin{tabular}{|l|ll|ll|}
 \hline
            &\multicolumn{2}{c|}{electron$+$jets} &
\multicolumn{2}{c|}{muon+jets}\\
            & $N_{\mathrm{jets}}= 3$ &  $N_{\mathrm{jets}}\geq 4$ &
$N_{\mathrm{jet}}= 3$ &  $N_{\mathrm{jets}}\geq 4$ \\
 \hline
 ${\rm t{\bar t}}$ & ~~157 $\pm$ 25  & 168 $\pm$ 27  & ~~197 $\pm$ 31  & 211 $\pm$ 33\\
 single top & ~~~~22 $\pm$  1 &~~~~9 $\pm$ 1 & ~~~~30 $\pm$  1 & ~~11 $\pm$  1\\
 W+jets     & ~~374 $\pm$ 27  & ~~94 $\pm$ 7  & ~~486 $\pm$ 34  & 115 $\pm$ 9\\
 Z+jets     & ~~~~66 $\pm$  5 & ~~15 $\pm$  1 & ~~~~46 $\pm$  3 & ~~11 $\pm$  1\\
 QCD        & ~~314 $\pm$ 19  & ~~53 $\pm$ 8  & ~~~~49 $\pm$  3 & ~~~~9 $\pm$ 1\\
 \hline
 sum (simulated events)     & ~~934 $\pm$ 55  & 339 $\pm$ 32   & ~~807 $\pm$ 53 & 358 $\pm$ 37\\
 \hline
 observed in data       & 1183            & 428            & 1064           & 423 \\
 \hline
     \end{tabular}
      \end{center}
\end{minipage}
\end{center}
\end{table*}

\section{Cross-Section Measurement}
\label{sec:xsec}
\subsection{Analysis Method}
\label{sec:KIT_EleJets}

We measure the ${\rm t{\bar t}}$ cross section in a data sample
consisting of events with leptons and jets in the final state. 
Since the event yields for the background processes can be difficult
to estimate purely from simulations,
the kinematic properties of the events are
used to separate signal from background. 
It would be natural to require four or more jets for the
selection of ${\rm t{\bar t}}$ events in the lepton$+$jets channel,
owing to the four final-state quarks present in these decays and 
because the number of background events from W/Z$+$jets and QCD
multijet events decreases with increasing jet multiplicity. However, it is
also useful to include events with only three
jets in the selection. In addition to improving the overall signal
efficiency, the inclusion of three-jet
events constrains the QCD and W$+$jet background normalization when a
simultaneous fit to
the three and four-or-more jet samples is performed.

The predicted jet-multiplicity distribution and the predicted ratio
between events
with three jets and events with four or more jets for the different
processes are used to simultaneously fit the fraction of ${\rm t{\bar t}}$
events and the contamination from background processes.
Kinematic
variables whose shapes are different for the different
processes are used to separate the backgrounds from the 
signal. 
After a number of 
variables and combinations were tested,
the variable M3 was chosen to
separate ${\rm t{\bar t}}$ events from background events
in the four-or-more-jet sample. 
This variable is defined as the
invariant mass of the combination of the three jets with the largest
vectorially summed transverse momentum. It approximates the mass of
the hadronically-decaying top quark and thus
provides good separation power. The three-jet sample is dominated by
W$+$jets events and QCD multijet events. For the three-jet sample,
a variable that is well suited for the discrimination of
QCD multijet events from the other processes is needed. 
In contrast to processes with
W bosons, QCD processes exhibit only small amounts
of missing transverse energy, mostly because of mismeasured jets
rather than the presence of neutrinos. Therefore \met~was chosen
as the discriminating variable to separate QCD events from W$+$jets
and ${\rm t{\bar t}}$ signal events in the three-jet sample. 

The \met~and M3 distributions from the observed data sample are fit
simultaneously to obtain the contributions of the signal and the main
background processes. We use a binned likelihood fit, where the number 
of expected events $\mu_j[i]$ in each bin $i$ of the distribution
for the variable of choice $j$ (either \met~or M3) is compared to the
observed number of events in this bin. 
The number of expected events in bin $i$ is given by:

\begin{equation}
\mu_{j}[i] = \sum_{k}\beta_{k}\cdot \alpha_{jk}[i]~,
\label{eq:mu_i}
\end{equation} 
where $\alpha_{jk}$ is the 
binned contribution (called ``template'' in the
following) for variable $j$ and process $k$.
The fit
parameters $\beta_{k}$ 
are the ratio of the 
measured ($\sigma_{k}$) and 
predicted ($\sigma_{k}^{\mathrm{pred}}$) cross sections for process $k$:

\begin{equation}
\beta_{k} = \frac{\sigma_{k}}{\sigma_{k}^{\mathrm{pred}}}.
\label{eq:beta_k_2}
\end{equation}
Here, $k$ denotes all processes
that are taken into account,
namely ${\rm t{\bar t}}$, W$+$jets, Z$+$jets,
single-top-quark decays, and QCD.
Since negative $\beta_k$ values
are unphysical, we do not
allow $\beta_k$ to become
smaller than zero in the fit. The templates $\alpha_{jk}$ are normalized to
the corresponding predictions for process $k$ for events with
three jets and events with four or more jets.
Because of the normalization of the fit templates to the prediction,
the fitted value of $\beta_{k}$ can be directly interpreted as
the scale factor one has to apply to the predicted
cross section of a given process $k$ to derive the measured cross section.

Templates for the ${\rm t{\bar t}}$, single-top,
W$+$jets, and Z$+$jets processes are derived from the
simulation, while the QCD multijet template is derived from data,
using a method that will be described later.
The shapes of the M3 
and \met~distributions of single-top-quark events are
very similar to the distributions of ${\rm t\bar{t}}$ events. Because of
this similarity and because of the very small number of expected
single-top-quark events, an unconstrained fit of the single-top-quark
contribution is not possible. However, since the single-top-quark
production process is theoretically well understood, the number of
such events
can be estimated from simulations. Therefore the
fit parameter for single top is not left to float freely,
but is instead subject to a Gaussian constraint with a mean of 1.0 and
width of 0.3.  The
uncertainty on this value is assigned according to the expected
precision of initial single-top cross-section measurements in
CMS~\cite{SingleTop10008}.
In addition, the ratio between the
W$+$jets and Z$+$jets cross sections 
is constrained to be within 30\% of the expectation
from theory, where the constraint width is set by the uncertainty in the NLO
cross sections~\cite{EWK10012}. These constraints are inserted by
multiplying the likelihood function used in the fit by Gaussian terms
of mean value 1.0 and widths corresponding to the uncertainties on the
respective constraints. The same $\beta_{k}$ parameters are used for both
jet-multiplicity bins.

A Neyman construction~\cite{Neyman} with central
intervals and a maximum-likelihood estimate of the ${\rm t{\bar t}}$
cross section as test statistic was chosen to obtain the confidence
interval for the ${\rm t{\bar t}}$ cross section. 
For this purpose pseudo-experiments are performed in which
the number of events from the different processes 
are chosen randomly around the values predicted by simulations 
within appropriate uncertainties. 
Specifically this is done by randomly choosing, for each background
process, an input value for $\beta_{k}$ from a normal distribution with mean
value 1.0 and a width of 30\% for W$+$jets, Z$+$jets, and single
top. Since the properties of QCD multijet events are more difficult
to calculate, a more conservative uncertainty of 50\% is used for this
background. The templates for the different processes are then scaled
with the corresponding $\beta_{k}$ values and summed together,
generating a pseudo-data distribution for \met~in the three-jet bin
and a pseudo-data distribution for M3 in the inclusive four-jet
bin. To simulate statistical fluctuations we then vary the contents of
each bin of the two distributions using Poisson statistics.
A maximum-likelihood fit to the templates is then performed on the
distributions. This procedure yields one signal fit result $\beta_{\rm t{\bar
t}}^{\mathrm{fit}}$ for each pseudo-experiment and provides a measurement of
any possible bias between the input and fitted values.
We vary the input value $\beta_{\rm t{\bar t}}^{\mathrm{in}}$ between
0.0 and 3.0 in steps of 0.2, and perform 50\,000 pseudo-experiments
for each value. Each set of pseudo-experiments gives a distribution of
the fitted values $\beta_{\rm t{\bar t}}^{\mathrm{fit}}$. 
For each input value we determine the median and the 68\% and 95\% quantiles 
of the corresponding $\beta_{{\rm t{\bar t}}}^{\mathrm{fit}}$ distribution and use 
these values for the estimation of the central values and for the construction of 
confidence belts, respectively. From this confidence-belt construction the \ttbar 
cross section result corresponding to the $\beta_{\rm t{\bar t}}^{\mathrm{fit}}$ measured 
in data can be extracted together with its total uncertainty.
By construction, this treatment the correct coverage probability.

\subsection{Systematic Uncertainties}

In general, the presence of a systematic uncertainty affects both the
number of selected events and the shape of the investigated
discriminating observables, resulting in modified distributions $\alpha^{\rm{syst}}_{jk}$ for the different
processes. In order to estimate the effect of different sources of
systematic uncertainties we construct modified templates and draw the
pseudo-data used for the statistical calculation from them. For each
source of systematic uncertainty $u$, two templates
$\alpha^{u,+1}_{jk}$ and
$\alpha^{u,-1}_{jk}$, corresponding to variations of $\pm1$ standard deviation
($\pm1\sigma$) of the specific systematic uncertainty, are
used. Both templates are normalized to the altered event yields for
each specific systematic uncertainty $u$, and thus account for both
changes in event rates and changes in parameter distributions. These
$\pm 1\sigma$
templates are derived either by
altering the nominal samples as described in the following sections or
from dedicated simulations.
The modified $\alpha^{u}_{jk}$ used for drawing 
the pseudo-data can then 
be constructed from the nominal template $\alpha_{jk}$
and the $\alpha^{u,\pm1}_{jk}$ templates. Therefore for
each uncertainty $u$ a strength parameter $\delta_{u}$
is introduced, and $\alpha^{\rm{syst}}_{jk}$ is defined as a linear
interpolation:

\begin{equation}
 \alpha^{\rm{syst}}_{jk}[i] =
\alpha_{jk}[i] + \sum_u |\delta_{u}|
\cdot (\alpha^{u,\mathrm{sign}(\delta_{u})}_{jk}[i]
- \alpha_{jk}[i])~.
 \label{eqn:syst_interpolation}
\end{equation}
Here, $u$ runs over all sources of systematic uncertainties and
$\alpha^{u,\pm1}_{jk}[i]$ is the prediction for bin $i$ of
distribution $j$ of process $k$ affected by $+1\sigma$ or $-1\sigma$
of uncertainty $u$. Random
numbers following a Gaussian distribution with a mean of zero and unit width
are used as values for each $\delta_{u}$, with the
strengths determined separately for each pseudo-experiment. The
nominal template is reproduced for
$\delta_{u} = 0$, while the two altered templates
correspond to $\delta_{u} = +1$ and
$\delta_{u} = -1$.
For all other values of $\delta_{u}$, the desired mixture of the
nominal and shifted templates is obtained.

In order to prevent unphysical negative predictions for a process, the
linear interpolation is cut off at a bin content of zero,
i.e., whenever a bin of 
$\alpha^{\rm{syst}}_{jk}$
calculated according to Eq. \ref{eqn:syst_interpolation} has a
value less than zero, zero is used instead. The definition of
$\beta_{\rm t{\bar t}}^{\rm{fit}}$ remains unchanged, meaning that the
original templates without uncertainties are employed for fitting the
pseudo-data distributions. 

The influence on our measurement due to the
imperfect knowledge 
of the jet energy scale (JES) 
is estimated by simultaneously 
varying all jet four-momenta either by $+1\sigma$ or by $-1\sigma$ of the
energy scale uncertainties, 
which are functions of the jet $\eta$ and $p_{\mathrm{T}}$. 
These uncertainties on the particle-flow 
jet energy scale are typically about 3\%, as shown in Ref.~\cite{JME-10-010}.
In addition, a constant 1.5\%
uncertainty due to changes in calorimeter calibrations and a
$p_{\mathrm{T}}$-dependent uncertainty 
of $1.32~\mathrm{GeV}/p_{\mathrm{T}}$ 
due to multiple collisions in the 
same event (``pileup'') are applied.
For jets that can be matched to a b~quark at the parton
level, we assign an additional uncertainty of 2\%\ if the jet lies within $|\eta| < 2.0$ and has
$p_{\mathrm{T}}$ between 50~GeV and
200~GeV, and 
a 3\%\ uncertainty otherwise. This uncertainty accounts for observed
response differences for b~jets generated in
\textsc{Pythia} and those in \textsc{Herwig}~\cite{Herwig}.
The overall uncertainty is determined by adding all of the individual
uncertainties in quadrature.

Jet asymmetry
measurements suggest that the jet energy resolutions (JER)
in data
are about 10\% worse than in the simulation~\cite{JME-10-014}. The uncertainty on
this measurement is also about 10\%. To account for this,
jets in the simulated samples are altered so that their resolutions match those measured in data, and
the effect is propagated to the calculation of \met. The impact of this
uncertainty on our measurement is determined by evaluating the change
in cross section when simulated jet resolutions are widened by 0\% or 20\%, rather than the default 10\%. 

The corrections in jet energy scale and resolution described above are
used to vary the missing transverse energy according to variations in
clustered jet energy. In order to also account for the uncertainty of
unclustered energy in \met, the amount of unclustered energy
contributing to \met~is shifted by $\pm 10\%$. However, the impact of
 the variation of the unclustered energy on the measurement is found to be negligible.

Adjusted simulated samples are used to evaluate the dominant
systematic uncertainites in the cross section measurements.
Two simulated samples of ${\rm t{\bar t}}$ events are
available to estimate the systematic uncertainty induced by the
lack of accurate knowledge of the amount of QCD initial-state (ISR) and final-state (FSR) radiation. For
one of these samples, the amount of ISR and FSR
has been increased, while less ISR and FSR is
assumed in the other sample.

The impact of uncertainties in the factorization scales on the
cross section measurement is estimated by varying the scales in each
of the samples by factors of 0.25 and 4.0 with respect to their
default values. The W$+$jets and Z$+$jets processes are treated as
being correlated, and their respective factorization scales are
shifted either down or up simultaneously, while the ${\rm t{\bar t}}$ scale is
considered to be uncorrelated and is shifted independently. The impact
of a variation of
the shower matching threshold is investigated by varying the matching thresholds for
the three processes by factors of 0.5 and 2.0
compared to the default thresholds. Again, W$+$jets and Z$+$jets processes are treated as
fully correlated and are varied simultaneously, while the ${\rm t{\bar
t}}$ process is considered uncorrelated and therefore independently altered.

The measurement of the electron $E_{\mathrm{T}}$ has a relative
uncertainty of 2.5\%\ in the endcap region, while the uncertainties
for the barrel region and for muons can be neglected. These variations
in the electron energy scale are also
propagated to the missing transverse energy. This 
component of the \met~uncertainty is
treated as uncorrelated with the other \met uncertainties. 

Correction factors have been applied to match the trigger-selection and lepton-selection 
efficiencies in simulated samples with those in data. These factors are
obtained from data by using decays of Z~bosons to dileptons.
In the electron$+$jets
channel, the correction factor is $0.933\pm0.032$.
In the muon$+$jets channel, it is $0.965 \pm 0.004$.
The uncertainty
on the ${\rm t{\bar t}}$ cross section measurement due to the
uncertainties of the correction factors is evaluated by weighting
all simulation-based samples according to the $\pm 1 \sigma$
uncertainties obtained from these studies. 

We evaluated the systematic uncertainty on the ${\rm t{\bar t}}$~cross section
measurement induced by the imperfect knowledge of the PDF of the
colliding protons using the
CTEQ6.6~\cite{cteq66} PDF set and the LHAPDF~\cite{lhapdf}
package. For this purpose, a reweighting procedure is applied to all
generated samples, in which each CTEQ6.6~PDF parameter is
independently varied by its positive and negative uncertainties, with
a new weight assigned to each variation.  The resulting templates are
used to estimate the impact of variations in the PDFs on our
measurement.

The default samples used for this analysis
were produced without any pileup collisions. These samples
are insufficient for the simulation of data taken in late 2010, when the
instantaneous luminosity was substantially higher than in early
data-taking and roughly four to five additional collisions per bunch
crossing were expected. 
In order to estimate the effect of these pileup collisions,
which are present in data but not in our fit templates, additional
samples of ${\rm t{\bar t}}$ and W$+$jets events that included the
simulation of these extra collisions were produced. Although the average
number of pileup collisions in these samples is slightly larger than the
expected number in data, these simulated events can still be used to provide a
conservative estimation of the impact of pileup collisions on our
measurement.

\subsection{Electron+Jets Analysis}

\begin{figure}[h!]
  \centering
    \includegraphics[width=0.49\textwidth]{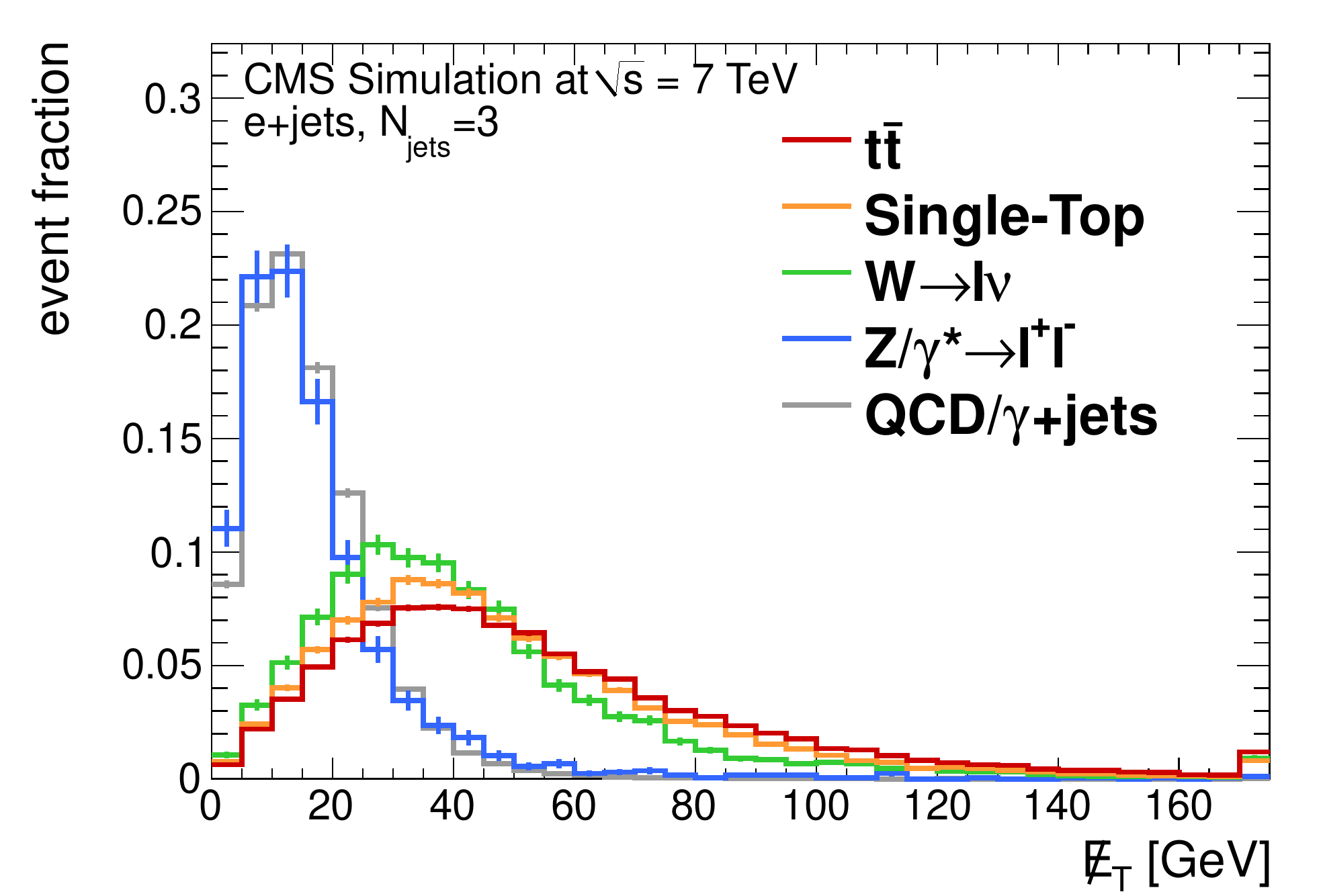}
     \includegraphics[width=0.49\textwidth]{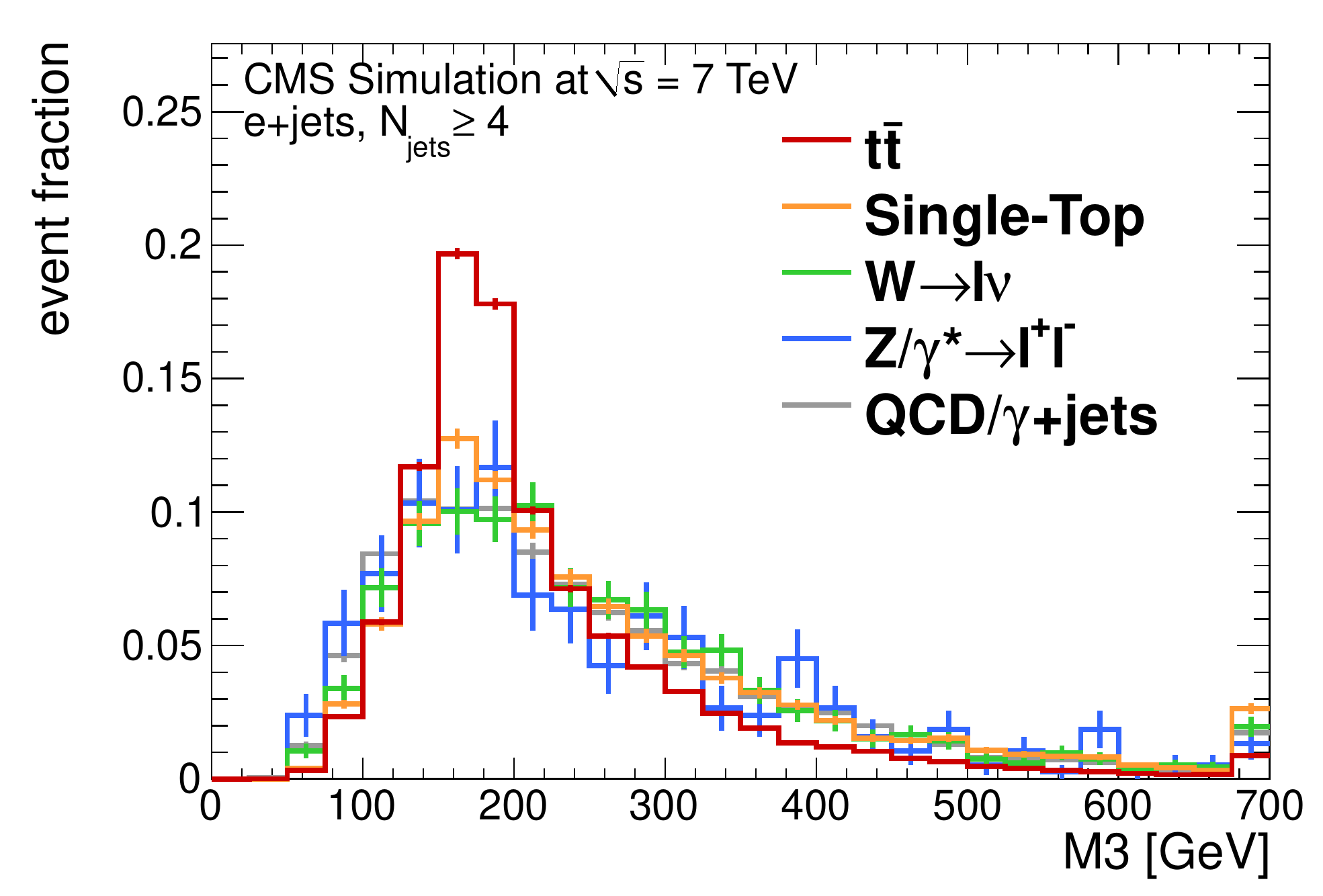}
    \caption{Simulated distributions from electron$+$jets events of
    (left)~\met~for events with three jets and (right)~M3 for events
    with four or more jets.  The contributions from the
    different processes are shown separately, and are normalized
    to unity. Error bars are statistical only.}
  \label{fig:KIT_Electron_Variables}
\end{figure}

In the electron$+$jets channel, we model ${\rm t{\bar t}}$,
W$+$jets, Z$+$jets, and single-top-quark production using
the simulated event samples
described previously.
For the QCD multijet background,
a sideband method based on data
is used to model the M3 and \met~distributions, where
the sidebands are chosen
to be depleted in
contributions from real W~bosons.
In the sideband selection, events must have an
electron that fails at least two of the three quality requirements:
$I_{\rm{rel}}<0.1$ (but $I_{\rm{rel}}<0.5$ is always required),
transverse impact parameter $<0.02$~cm, and the standard 
electron identification criteria.
As verified with simulated events, the data sample extracted in
this way has a QCD
multijet purity larger than 99\%. In addition, the M3 and
\met~shapes derived from this sample are in good agreement with the
distributions from the simulation.
The fraction of events in the three-jet and inclusive-four-jet sample
for each process are taken from the simulation.
Figure~\ref{fig:KIT_Electron_Variables} shows the
distributions of \met and M3 from the simulated three-jet and
four-or-more jet samples, respectively, for the different processes.

The modeling of QCD~multijet events from data might induce an additional
source of systematic uncertainty.  This is investigated by
separating the sideband region from which the QCD templates are
derived into two parts. The sideband region is defined, in addition
to other criteria, by $0.1 < I_{\mathrm{rel}} < 0.5$ for the electron$+$jets
channel.
The QCD template is further split into two separate
samples of equal-width regions in $I_{\mathrm{rel}}$
($0.1 < I_{\mathrm{rel}} < 0.3$ and $0.3 < I_{\mathrm{rel}} < 0.5$),
and the templates
from these two samples are 
used to estimate this systematic uncertainty.
The uncertainty in the ratio of the number of events with three jets
to that with four or more jets is investigated
as well. While this ratio for
the model predictions is taken from the simulation, the observed ratio
in the sideband selection is different. Consequently, the two
potential sources of systematic uncertainties are studied
separately via two independent strength parameters.

The ${\rm t{\bar t}}$~cross section is measured, accounting for statistical
and systematic uncertainties, using the fit method described in
Section~\ref{sec:KIT_EleJets}.  The parameter $\beta_{\rm t{\bar
t}}^{\textrm{fit}}$, which is used to compute the ${\rm t{\bar
t}}$~cross  section, and the values of $\beta_k$ for the background
processes are determined in the fit.
The results for $\beta_{\rm t{\bar t}}$ and the signal and background
event yields for the inclusive
three-jet bin are given in Table~\ref{tab:Electron_DataResults}. While
the number of fitted ${\rm t{\bar t}}$
events N$_{\rm t{\bar t}}$ is quoted with its combined statistical and
systematic uncertainty, for
the remaining processes only statistical uncertainties are given. A
list of all systematic uncertainties in this channel is provided in
Section~\ref{sec:Combination}, with the dominant systematic
uncertainty coming from the lack of knowledge of the jet energy
scale.

In the
electron$+$jets channel, the resulting
$\rm t{\bar t}$~production cross section is:

\begin{equation}
 \sigma_{\rm t{\bar t}} = 180^{+45}_{-38}\,\mathrm{(stat. + syst.)}
  \pm 7\,\mathrm{(lumi.)}\,\mathrm{pb\, .}
\end{equation}
The fit produces a combined statistical and systematic uncertainty, as
given above.  A fit using only the nominal templates yields a
statistical uncertainty of $^{+23}_{-22}\,\mathrm{pb}$.
Assuming uncorrelated, Gaussian behaviour of
statistical and systematic uncertainties, one can subtract
the statistical uncertainty in quadrature from the overall
uncertainty, resulting in a systematic uncertainty of
$^{+39}_{-31}\,\mathrm{pb}$.  Individual uncertainties are summarized
in Section~\ref{sec:Combination}.

\begin{table*}[h!]
  \begin{center}
      \caption{The predicted and fitted values for $\beta_{\rm t{\bar t}}$ and for the numbers
      of events for the various contributions from the inclusive three-jet electron$+$jets
      sample. The quoted
      uncertainties in the $\rm{ t{\bar t}}$ yield account for statistical and
      systematic uncertainties, while the uncertainties in the background
      event yields are derived from the covariance matrix of the
      maximum-likelihood fit and therefore represent purely
      statistical uncertainties.}
\setlength{\extrarowheight}{2.5pt}

        \begin{tabular}{|c|c|c|c|c|c|c|}\hline
                             &  $\beta_{\rm t{\bar t}}$ & N$_{\rm t{\bar t}}$   & N$_{\textrm{single-top}}$ &  N$_{W+\textrm{jets}}$  &  N$_{Z+\textrm{jets}}$  &  N$_{\textrm{QCD}}$  \\ \hline
 electron$+$jets (predicted) &       $1.00$             & $325\pm 52$           & $31\pm 2$                &  $468\pm 34$            &  $81\pm 6$              &  $367\pm 27$ \\ \hline
 electron$+$jets (fitted)    &       $1.14^{+0.29}_{-0.24}$             & $371^{+94}_{-78}$     & $33\pm 9$                &  $669\pm 61$            &  $116\pm 36$            &  $422\pm 51$ \\ \hline
    \end{tabular}
    \label{tab:Electron_DataResults}
  \end{center}
\end{table*}

\begin{figure}[h!]
  \centering
  \includegraphics[angle = 0, width=0.49\textwidth]{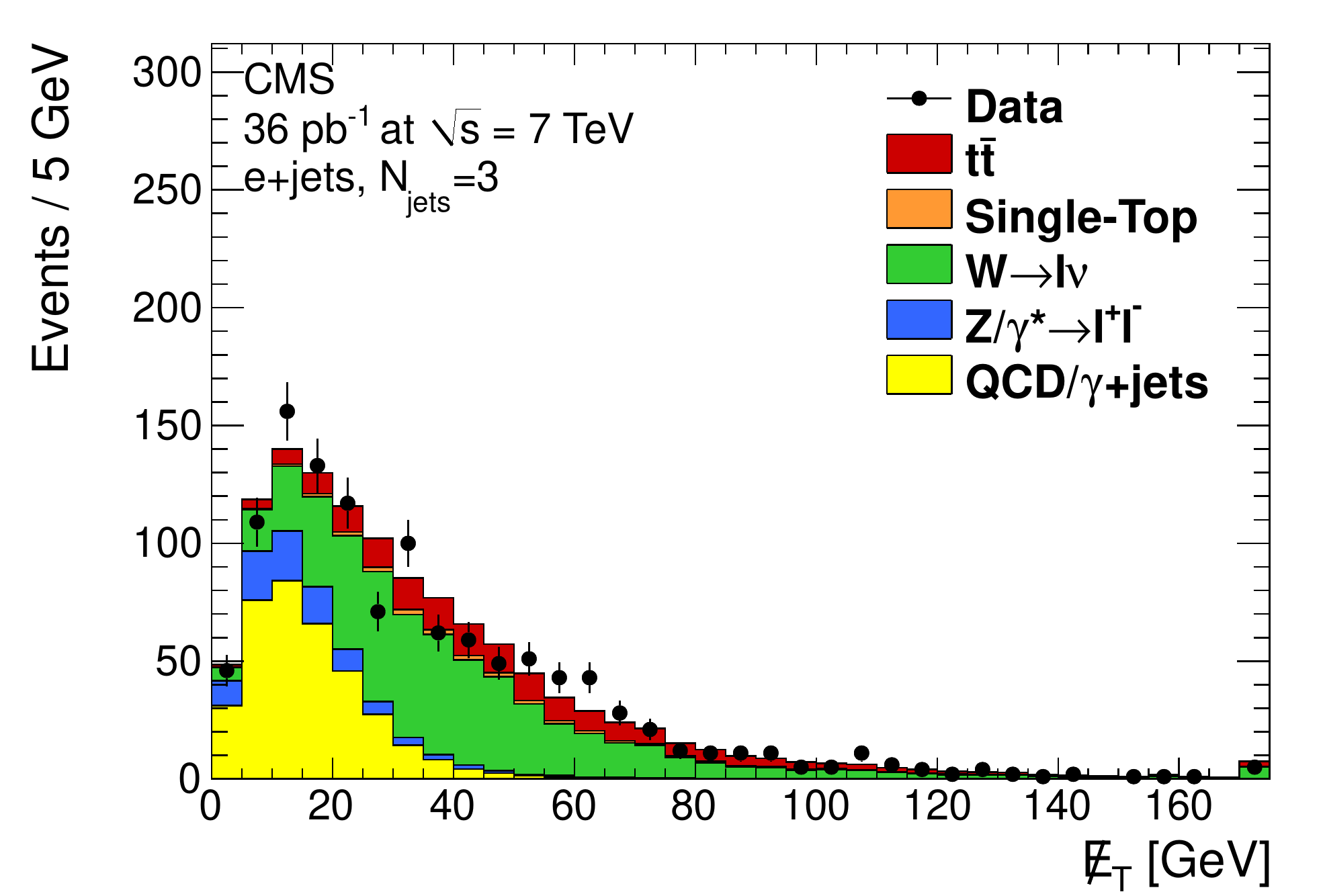}
  \includegraphics[angle = 0, width=0.49\textwidth]{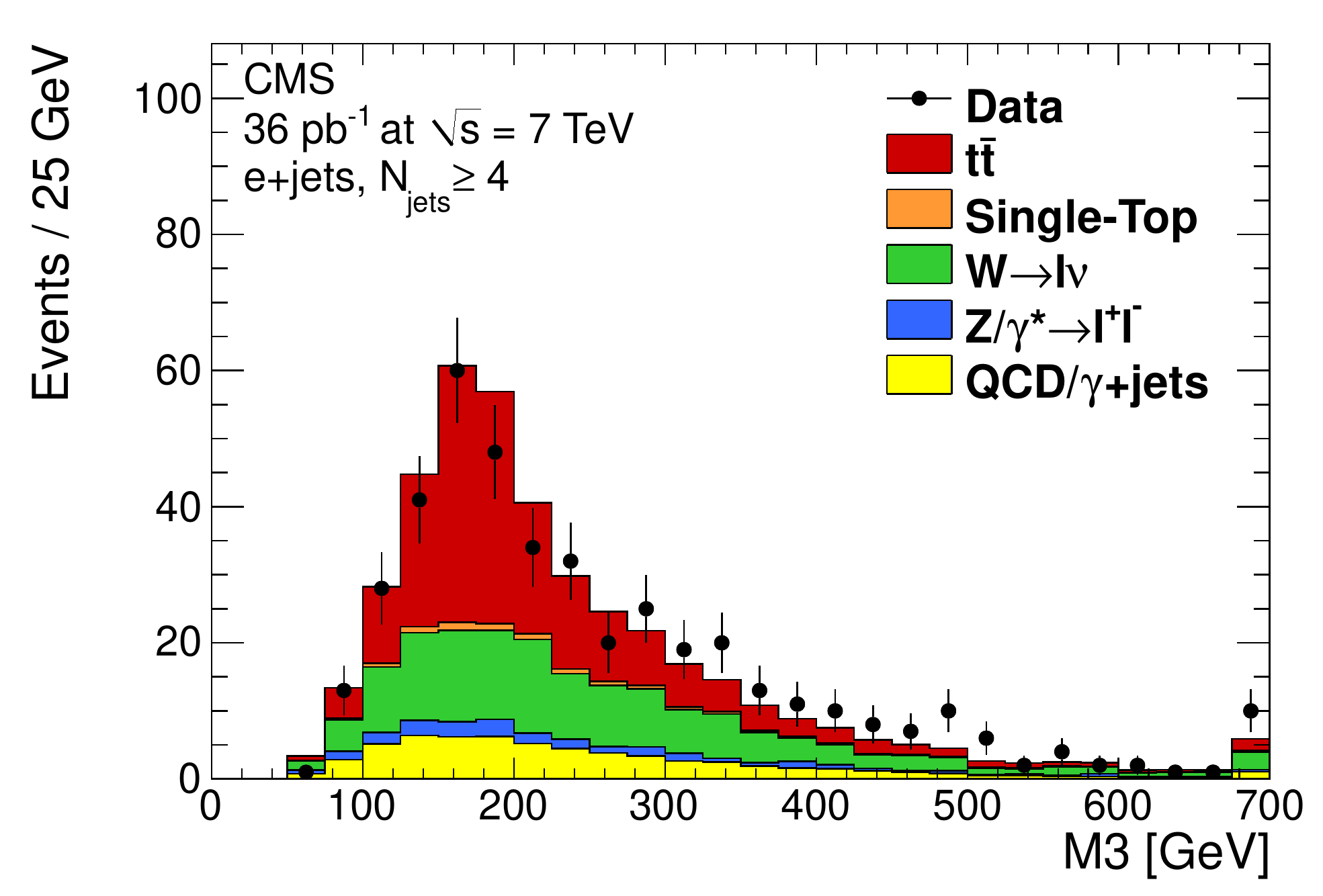}
 \caption{Electron$+$jets channel: Comparison of the
distributions in data and simulation of the discriminating variables
  \met~(left) and
M3 (right) for signal and background.
The simulation has been
normalized to the fit results. Only statistical uncertainties are shown.}
 \label{fig:KIT_Electron_DataMC_MET_M3}
\end{figure}

The measured $\rm{ t{\bar t}}$~cross section, in
combination with the background estimation, can be used to compare
distributions of \met and M3 found in data with those
predicted by Monte Carlo simulation.
Figure~\ref{fig:KIT_Electron_DataMC_MET_M3} shows the distributions of
the missing transverse energy and M3 as observed in data. For
comparison, the templates from simulation
are normalized to the fitted fractions.
The deviation visible in the high-M3 region between simulation and data has
been investigated using pseudo-experiments including statistical and
systematic uncertainties. For 10\% of the simultaneous fits to \met~and M3 in
these pseudo-experiments, the derived Kolmogorov-Smirnov (KS) value is larger
than the KS value observed in data. Therefore, the observed deviation in the
M3 distribution is not outside the range of expected fluctuations.

\subsection{Muon+Jets Analysis}
\label{sec:muon_jets}
\begin{figure}[h!]
 \centering
   \includegraphics[angle =
   0,width=0.49\textwidth]{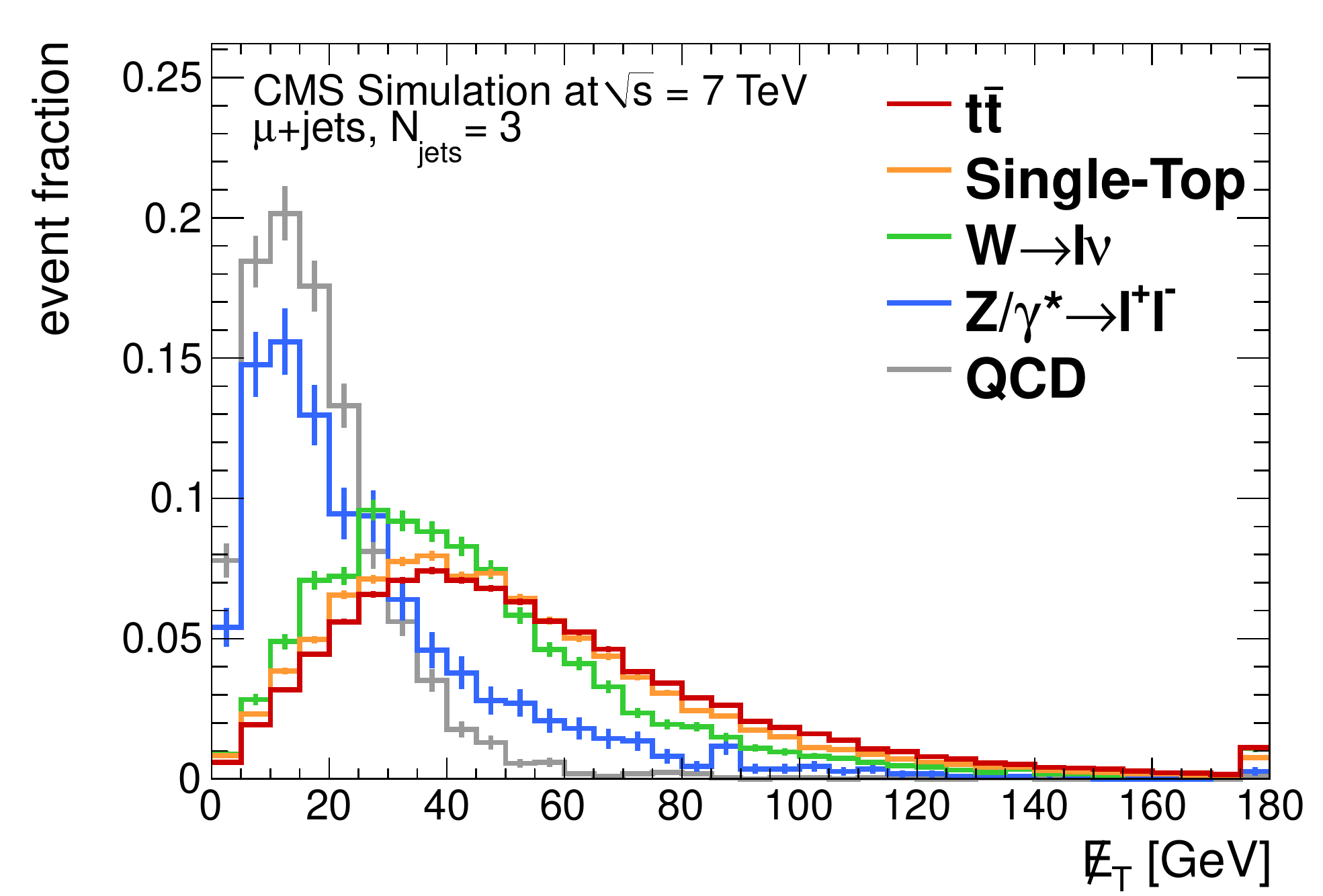}
   \includegraphics[angle =
   0,width=0.49\textwidth]{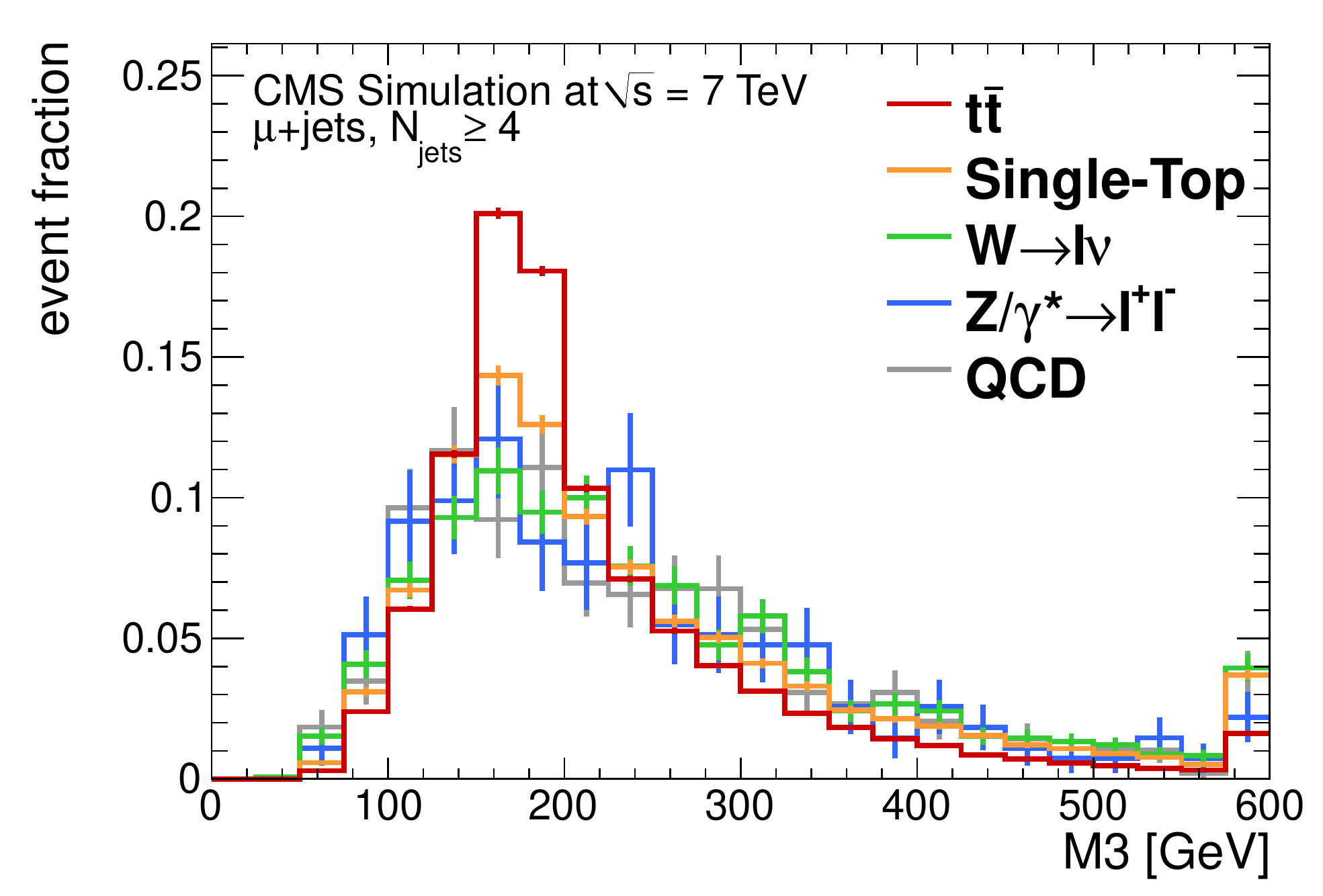}
   \caption{Simulated distributions from muon$+$jets events of
   (left)~\met for events with three jets and (right)~M3 for events
   with four or more jets for the different processes.  The
   contributions from the different processes are shown separately,
   and are normalized to unity. Error bars are statistical only.}
 \label{fig:KIT_MuonShapes}
\end{figure}

The same analysis method is
used to measure the
$\rm{ t{\bar t}}$~cross section in the
muon$+$jets final state. \met~and M3 are again used as discriminating variables. Shape comparisons for the
different physics processes are shown in
Fig.~\ref{fig:KIT_MuonShapes}.
In the muon$+$jets channel, the QCD templates for these two
distributions are derived from data by selecting events in a sideband
region enriched in QCD multijet events.
The relative isolation is required to be between
0.2 and 0.5 for the sideband selection, in contrast to the
nominal selection, where
the muon must have a relative isolation smaller than 0.05.
The gap between the allowed isolation ranges in the two selections
reduces the signal events contribution to the sideband. Events
containing muons with large
relative-isolation values have different kinematics due to the
correlation of the relative isolation with transverse momentum.
We therefore restrict $I_{\mathrm{rel}}$ to be smaller than 0.5.
The QCD multijet
purity as measured from simulation is $98.4\%$ in the three-jet sample
and $94.3\%$ for events with four or more jets.
As in the electron$+$jets channel, the QCD template is split into two
separate samples of equal-width regions
in $I_{\mathrm{rel}}$, and the templates
from these two samples are used to estimate the systematic uncertainty on the QCD modeling.
Apart from the electron energy
scale, all other systematic uncertainties described in the previous
section are also accounted for in the muon$+$jets
analysis. A summary of the individual
contributions from the various sources of systematic uncertainties is
provided in Section~\ref{sec:Combination}.

\begin{figure}[h!]
  \centering
    \includegraphics[angle = 0, width=0.49\textwidth]{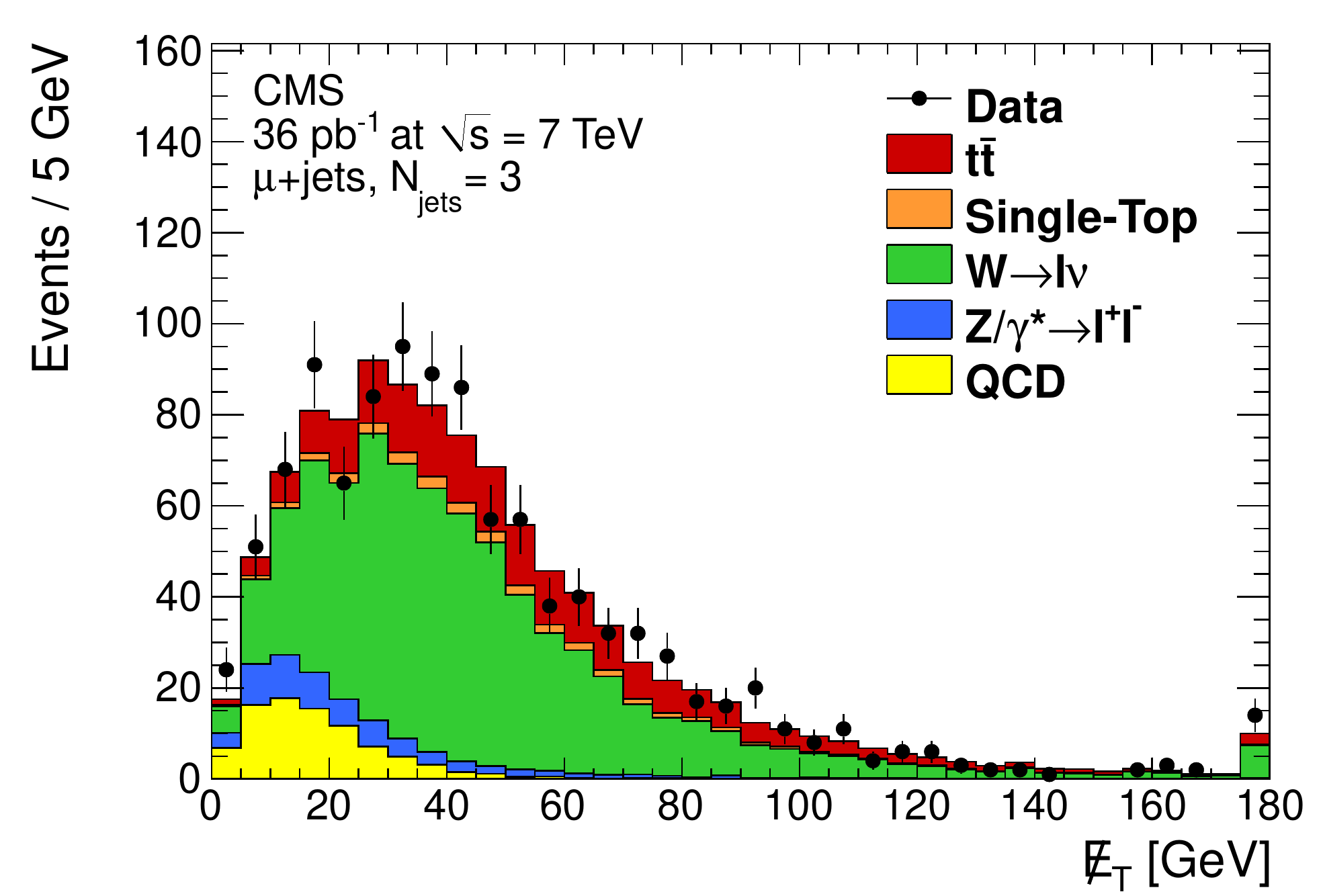}
    \includegraphics[angle = 0,
    width=0.49\textwidth]{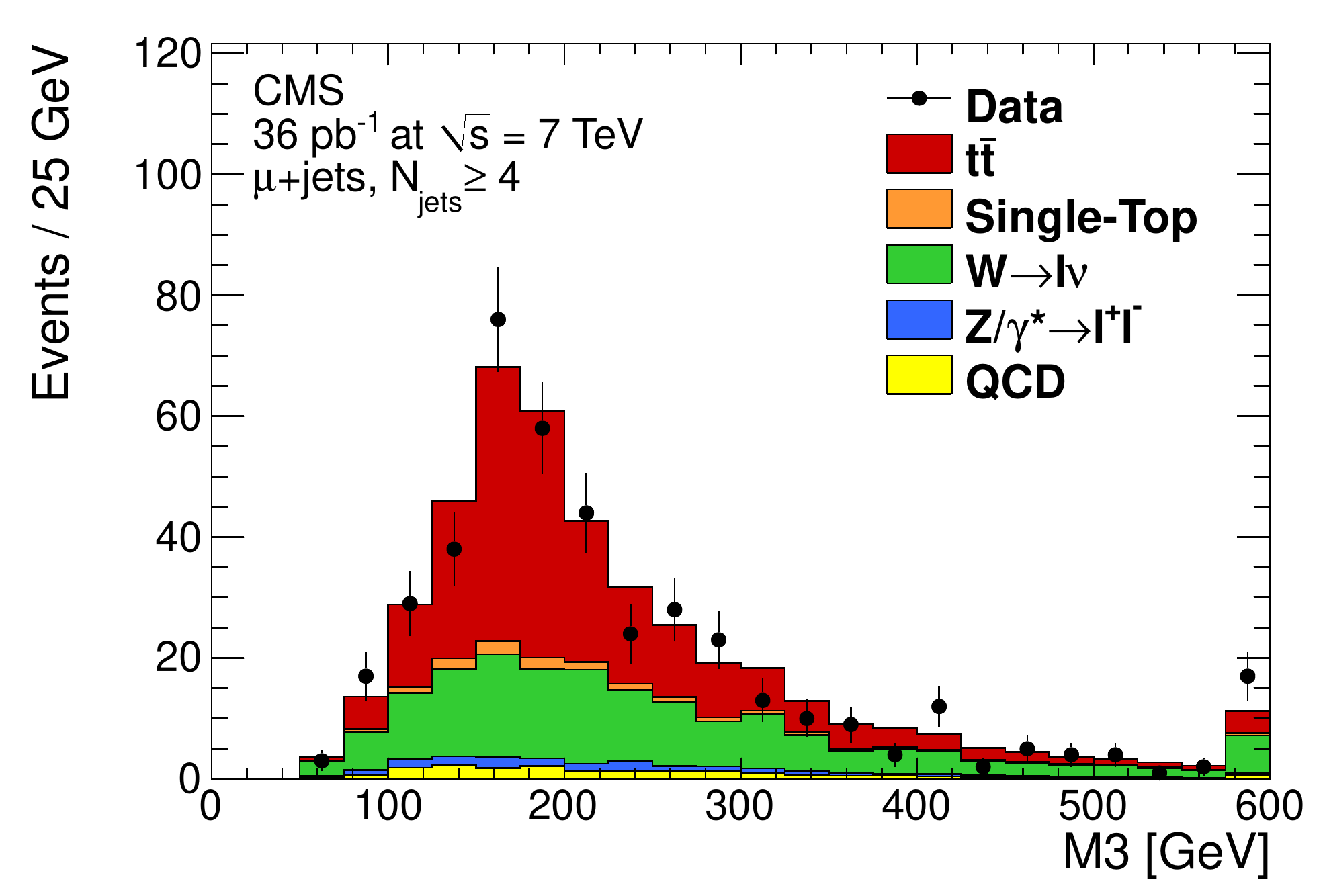}
   \caption{Muon$+$jets channel: Comparison of the data distributions of
    the discriminating variables \met~(left) and M3 (right) and the
    simulation of the different processes. The simulation has been
    normalized to the fit results.}
    \label{fig:Muon_MCScaledToFit}
\end{figure}

The results for $\beta_{\rm t{\bar t}}$ and the various background
yields from the binned likelihood fit to the inclusive three-jet
muon$+$jets sample are given in
Table~\ref{tab:Muon_DataResults}. 
Using the method described in Section~\ref{sec:KIT_EleJets}, the fitted
value $\beta_{\rm t{\bar t}}$ and the
$\pm1\sigma$ statistical$+$systematic~uncertainties corresponding to the
fitted value
$\beta_{\rm t{\bar t}}^{\textrm{fit}}=1.07$ are determined. The result
of the muon$+$jets analysis is a measured ${\rm t{\bar t}}$~production cross
section of:

\begin{equation}
 \sigma_{\rm t{\bar t}} = 168^{+42}_{-35}\,\mathrm{(stat. + syst.)}
\pm 7\,\mathrm{(lumi.)}\,\mathrm{pb\,.}
\end{equation}
The statistical uncertainty is
$^{+18}_{-17}\,\mathrm{pb}$. With the assumption that the statistical
and systematic
uncertainties are Gaussian and uncorrelated, the systematic
uncertainty is calculated to be
$^{+38}_{-31}\,\mathrm{pb}$. Similar to the measurement in the electron$+$jets channel, the
fitted numbers of W$+$jets, Z$+$jets events and QCD multijet events
are found to exceed the predicted values.
The KS $p$-value of this fit result has been determined to be 95\%.
Figure~\ref{fig:Muon_MCScaledToFit}
shows comparisons of the distributions of \met~and M3 between data
and simulation, where the simulation has been scaled to the results
obtained from the binned likelihood fit.

\begin{table*}[h!]
  \begin{center}
      \caption{The predicted and fitted values for $\beta_{\rm t{\bar t}}$ and for the numbers
      of events for the various contributions from the
      inclusive three-jet muon$+$jets sample. The quoted
      uncertainties in the $\rm{ t{\bar t}}$ yield account for statistical and
      systematic uncertainties, while the uncertainties in the background
      event yields are derived from the covariance matrix of the
      maximum-likelihood fit and
      therefore represent purely statistical uncertainties.}
\setlength{\extrarowheight}{2.5pt}

        \begin{tabular}{|c|c|c|c|c|c|c|}\hline
                         &  $\beta_{\rm t{\bar t}}$ & N$_{\rm t{\bar t}}$ &   N$_{\textrm{single-top}}$ &    N$_{W+\textrm{jets}}$  &  N$_{Z+\textrm{jets}}$  &    N$_{\textrm{QCD}}$  \\ \hline
 muon$+$jets (predicted) &       $1.00$             & $408\pm 64$         &   $41\pm 2$                &      $601\pm 43$          &  $57\pm 4$              &  $58\pm 4$ \\ \hline
 muon$+$jets(fitted)     &         $1.07^{+0.26}_{-0.24}$           & $437^{+106}_{-90}$  &   $41\pm 12$               &  $813\pm 59$              &  $76\pm 22$             &  $123\pm 33$ \\ \hline
    \end{tabular}
    \label{tab:Muon_DataResults}
  \end{center}
\end{table*}

\subsection{Combined Electron+Jets and Muon+Jets Analysis}
\label{sec:Combination}

\begin{figure}[h!]
  \centering
    \includegraphics[angle = 0,
    width=0.46\textwidth]{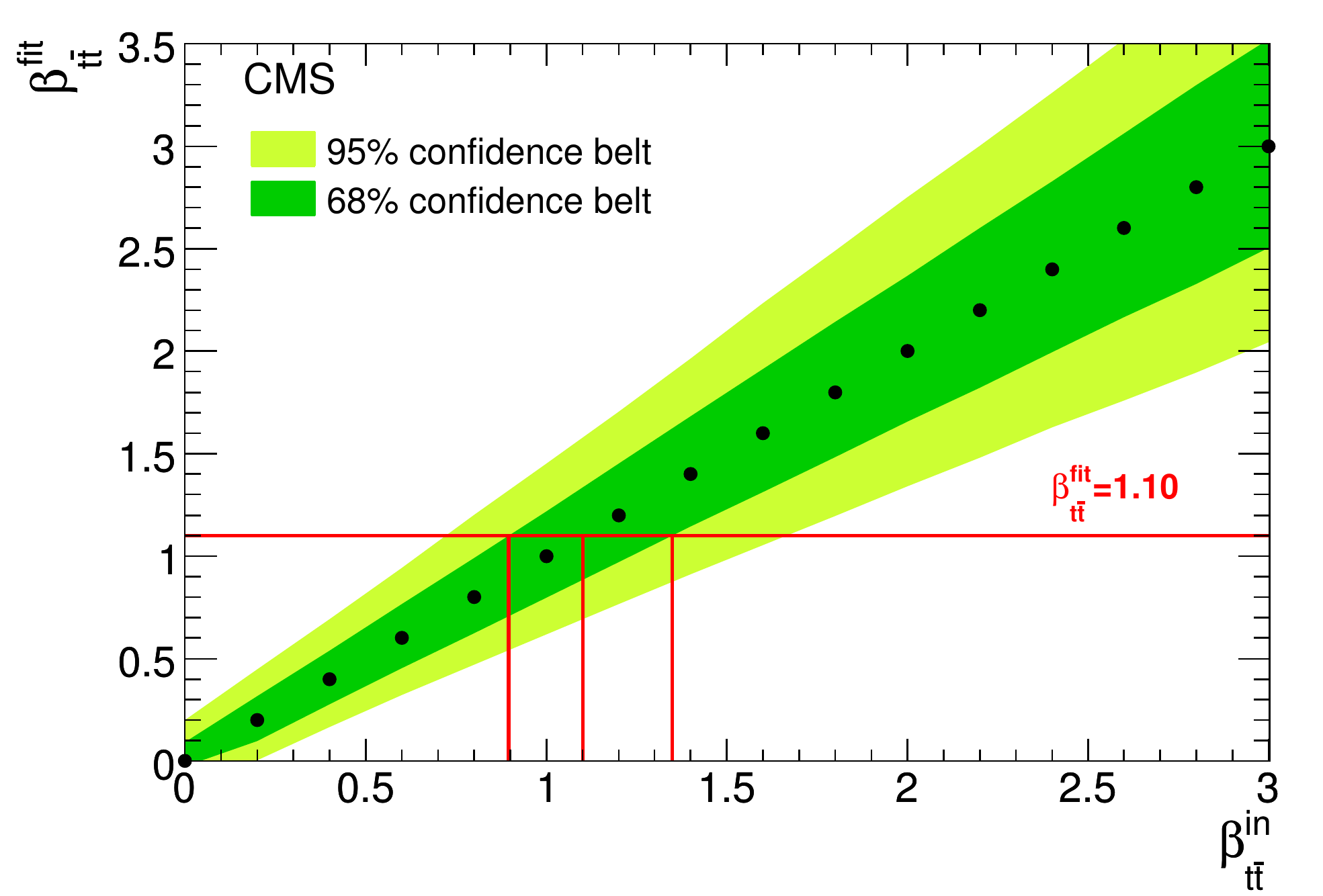}
    \caption{Neyman construction including all systematic
    uncertainties for the combined measurement of the ${\rm t{\bar
    t}}$~production cross section in the electron$+$jets and
    muon$+$jets channels. The horizontal line indicates the determined
    value $\beta_{\rm t{\bar t}}= 1.10$ from the binned likelihood fit
    to observed pp~collision data.
}
    \label{fig:Combination_FinalNCWithDataFit}
\end{figure}

The ${\rm t{\bar t}}$ cross section is also determined
using the method described in Section~\ref{sec:KIT_EleJets}
for the combined electron$+$jets and muon$+$jets channel.
Simultaneous fits of the \met~and M3 distributions are performed in
both the electron$+$jets and muon$+$jets channels.
Six fit parameters are used: the fraction
of ${\rm t{\bar t}}$~events ($\beta_{\rm t{\bar t}}$),
the fractions of the different
background processes ($\beta_{\rm t}$, $\beta_{\rm W}$, and
$\beta_{\rm Z}$), and two distinct fractions of QCD multijet events
($\beta_{\mathrm{QCD},e}$ and $\beta_{\mathrm{QCD},\mu}$). The use of
two fit parameters for the fraction of QCD~multijet~events is
motivated by the fact that the sources of such events contributing to
this background in the electron$+$jets channel are very different from
those contributing to the muon$+$jets channel. The cross section was
determined with the same procedure used for the individual electron and muon
channels. Figure~\ref{fig:Combination_FinalNCWithDataFit} shows the
Neyman construction with all systematic uncertainties included for the
combined measurement. The fitted $\beta_{\rm t{\bar t}}$ parameter and
the fitted numbers of events for the various background processes are
summarized in Table~\ref{tab:Combination_DataResults}.

\begin{table*}[h!]
  \begin{center}
      \caption{Predicted and fitted values for $\beta_{\rm t{\bar t}}$ and for the  numbers
      of events for the various contributions in the inclusive
      three-jet combined electron$+$jets and muon$+$jets sample.
      The quoted
      uncertainties in the $\rm{t{\bar t}}$ yield account for statistical and systematic
      uncertainties, while the uncertainties in the background event
      yields are derived from the covariance matrix of the
      maximum-likelihood fit and therefore represent purely statistical uncertainties.}
\setlength{\extrarowheight}{2.5pt}

        \begin{tabular}{|c|c|c|c|c|c|c|c|}\hline
               &  $\beta_{\rm t{\bar t}}$ & N$_{\rm t{\bar t}}$ & N$_{\textrm{single-top}}$ &  N$_{W+\textrm{jets}}$  &  N$_{Z+\textrm{jets}}$ &  N$_{\textrm{QCD}}$ e+jets & N$_{\textrm{QCD}}$ $\mu$+jets\\ \hline
     predicted &      $1.00$              & $733\pm 116$        & $72\pm 4$                &  $1069\pm 77$           &  $138\pm 10$           &  $367\pm 27$               & $58 \pm 4$ \\ \hline
        fitted &      $1.10^{+0.25}_{-0.21}$              & $806^{+183}_{-154}$ & $ 76\pm 22 $             &  $1475\pm 86$           &  $184\pm 51$           &  $440\pm 44$               &  $113\pm 31$                 \\ \hline
    \end{tabular}
    \label{tab:Combination_DataResults}
  \end{center}
\end{table*}

The fitted $\beta_{\rm t{\bar t}}$ value corresponds to a measured
$\rm t{\bar t}$ cross section in the lepton$+$jets channel of

\begin{equation}
 \sigma_{\rm t{\bar t}} =
173^{+39}_{-32}\,\mathrm{(stat. + syst.)}
\pm 7\,\mathrm{(lumi.)}\,\mathrm{pb}~.
\end{equation}
The statistical uncertainty is 14~pb. Subtracting this in quadrature
from the overall uncertainty yields a systematic uncertainty of
 $^{+36}_{-29}\,\mathrm{pb}$. The fit in the combined channel
yields a KS $p$-value of 68\% and agrees well with a simple average of
the results in the muon and electron channels, while correctly accounting for correlations.

Table~\ref{tab:All_Systematics} gives an overview of the
estimated statistical and systematic uncertainties for this combined
measurement as well as for the two channels separately. The different sources of systematic uncertainties are
treated as fully correlated between the two channels, except for
flavour-specific QCD and lepton uncertainties, which are assumed to be
uncorrelated. In order to estimate the impact
of individual systematic uncertainties, Neyman constructions
where only the specific source of systematic uncertainty under study
is accounted for are used. Each result indicates the
combined statistical and systematic
uncertainties of the contribution under study.

One can see that the largest contributor to the
overall systematic uncertainty is the uncertainty in the jet
energy scale. Combining both channels significantly reduces the
statistical uncertainty in the measured cross section. However, since
both single measurements are already dominated by systematic
uncertainties, the improvement in the total uncertainty of the combined
measurement is relatively small.

\begin{table*}[h!]
  \begin{center} \caption{Relative statistical and systematic
  uncertainties in the estimation of the ${\rm t{\bar t}}$ production
  cross section in the electron$+$jets and muon$+$jets channels, and
  their combination, assuming $\beta_{\rm t{\bar t}}=1$. The total (``stat.$+$syst.'') uncertainty is obtained from a Neyman construction,
  for which all sources of systematic uncertainties are taken into
  account in the prior predictive ensembles.  The estimate of
  each systematic uncertainty (``syst. only'') is calculated by assuming
  uncorrelated, Gaussian behaviour of the statistical and systematic
  uncertainties and subtracting the statistical uncertainty in
  quadrature from the total uncertainty.}
  {
\small

\setlength{\extrarowheight}{2.5pt}
   \begin{tabular}{|l|c|c|c|c|c|c|}\hline
      & \multicolumn{2}{c|}{electron$+$jets channel} &
      \multicolumn{2}{c|}{muon$+$jets
      channel}&\multicolumn{2}{c|}{combined result} \\
      \cline{2-7}
      & stat.+syst. & syst. & stat.+syst. & syst. & stat.+syst. & syst. \\
      & uncertainty & only & uncertainty & only & uncertainty & only \\ \hline
        Stat.~uncertainty           &  $_{-13.1\%}^{+14.0\%}$ & -- &
	$_{-10.8\%}^{+11.4\%}$& -- & $_{-8.4\%}^{+8.7\%}$ & --    \\
	\hline
        JES                         &  $_{-20.4\%}^{+23.5\%}$ &
	$_{-15.6\%}^{+18.9\%}$ & $_{-18.8\%}^{+21.9\%}$ &
	$_{-15.4\%}^{+18.7\%}$ & $_{-17.6\%}^{+20.3\%}$ &
	$_{-15.5\%}^{+18.3\%}$   \\[+0pt] \hline
	Factorization scale              &  $_{-14.3\%}^{+15.5\%}$ &
	$_{-5.7\%}^{+6.7\%} $&
	$_{-12.9\%}^{+13.8\%}$ & $_{-7.1\%}^{+7.8\%}$ &
	$_{-10.6\%}^{+11.2\%}$  &  $_{-6.5\%}^{+7.1\%}$ \\[+0pt]
	\hline
	Matching threshold               &  $_{-14.0\%}^{+15.0\%}$ &
	$_{-4.9\%}^{+5.4\%}$ &
	$_{-12.9\%}^{+14.1\%}$ & $_{-7.1\%}^{+8.3\%}$ &
	$_{-9.8\%}^{+10.5\%}$ & $_{-5.0\%}^{+5.9\%}$   \\[+0pt] \hline
	Pileup                           &  $_{-13.8\%}^{+14.4\%}$ &
	$_{-4.3\%}^{+3.4\%}$ &
	$_{-11.3\%}^{+11.7\%}$ & $_{-3.3\%}^{+2.6\%}$ &
	$_{-9.3\%}^{+9.3\%}$ & $_{-4.0\%}^{+3.3\%}$  \\[+0pt] \hline
	ID/reconstruction          &  $_{-13.6\%}^{+14.5\%}$ &
	$_{-3.7\%}^{+3.8\%}$ &
	$_{-11.2\%}^{+11.9\%}$ & $_{-3.0\%}^{+3.4\%}$ &
	$_{-8.7\%}^{+9.2\%}$ & $_{-2.3\%}^{+3.0\%}$   \\[+0pt] \hline
	QCD rate \& shape                &  $_{-14.8\%}^{+14.7\%}$ &
	$_{-6.9\%}^{+4.5\%}$ &
	$_{-10.9\%}^{+11.4\%}$ & $_{-1.5\%}^{+0.0\%}$ &
	$_{-8.9\%}^{+9.1\%}$ & $_{-2.9\%}^{+2.7\%}$   \\[+0pt] \hline
	ISR/FSR variation                &  $_{-13.3\%}^{+14.0\%}$ &
	$_{-2.3\%}^{+0.0\%}$ &
	$_{-11.3\%}^{+11.9\%}$ & $_{-3.3\%}^{+3.4\%}$ &
	$_{-8.6\%}^{+9.0\%}$ &  $_{-1.8\%}^{+2.3\%}$ \\[+0pt] \hline
        JER                              &  $_{-13.1\%}^{+14.0\%}$ &
	$_{-0.0\%}^{+0.0\%}$ & $_{-10.8\%}^{+11.4\%}$ &
	$_{-0.0\%}^{+0.0\%}$ & $_{-8.4\%}^{+8.8\%}$ &
	$_{-0.0\%}^{+1.3\%}$    \\[+0pt] \hline
	PDF uncertainty                  &  $_{-13.1\%}^{+14.0\%}$ &
	$_{-0.0\%}^{+0.0\%}$ &
	$_{-10.9\%}^{+11.4\%}$ & $_{-1.5\%}^{+0.0\%}$ &
	$_{-8.5\%}^{+8.7\%}$ &  $_{-1.3\%}^{+0.0\%}$  \\[+0pt] \hline
	\hline
        Total                            &  $_{-22.2\%}^{+26.6\%}$ &
	$_{-17.9\%}^{+22.6\%}$ &
	$_{-20.9\%}^{+25.3\%}$ & $_{-17.9\%}^{+22.6\%}$ &
	$_{-19.3\%}^{+23.5\%}$ & $_{-17.4\%}^{+21.8\%}$   \\[+0pt] \hline
    \end{tabular}
    \label{tab:All_Systematics}
    }
  \end{center}
\end{table*}

\begin{figure}[h!]
 \centering
   \includegraphics[angle = 0,
width=0.49\textwidth]{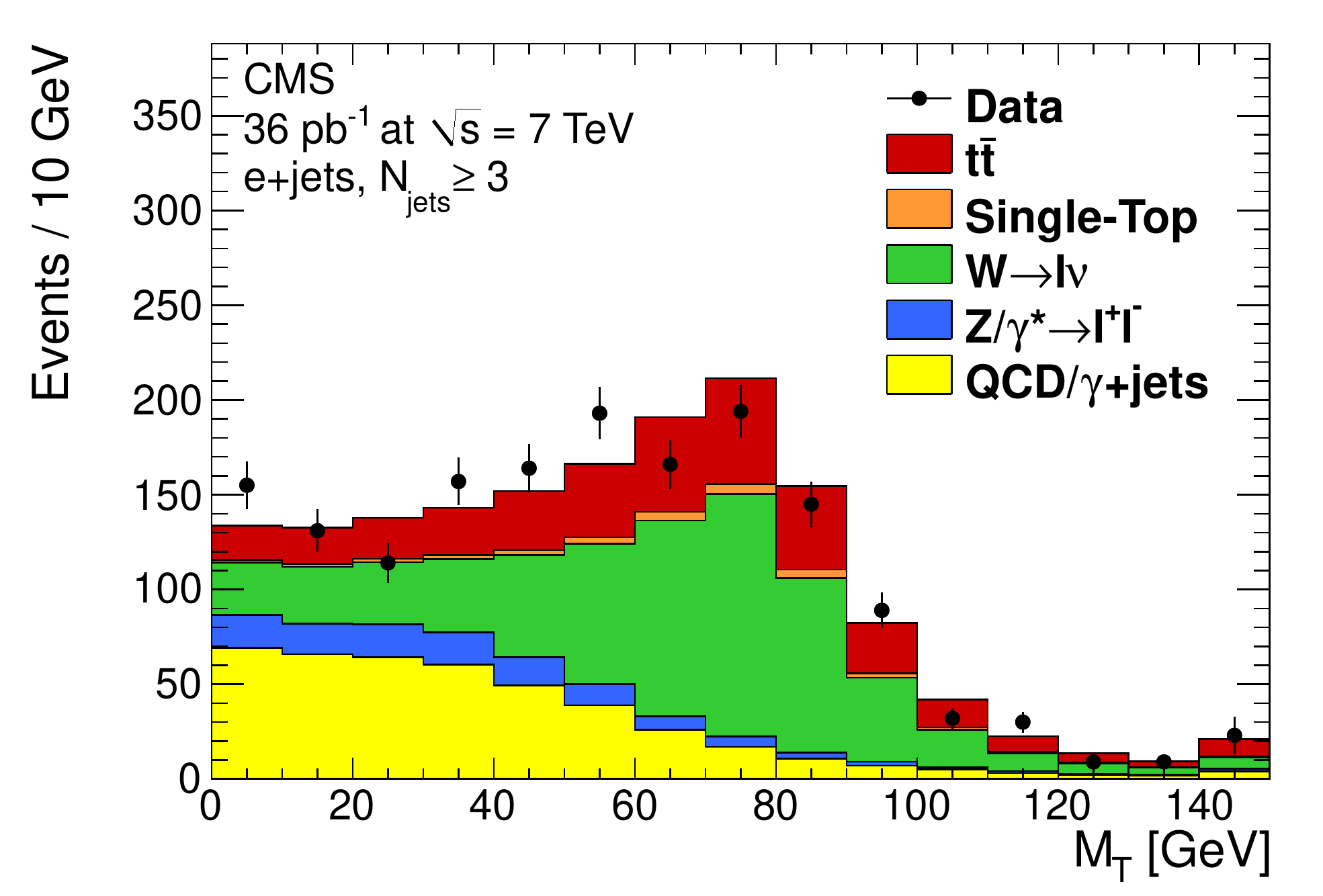}
   \includegraphics[angle = 0,
width=0.49\textwidth]{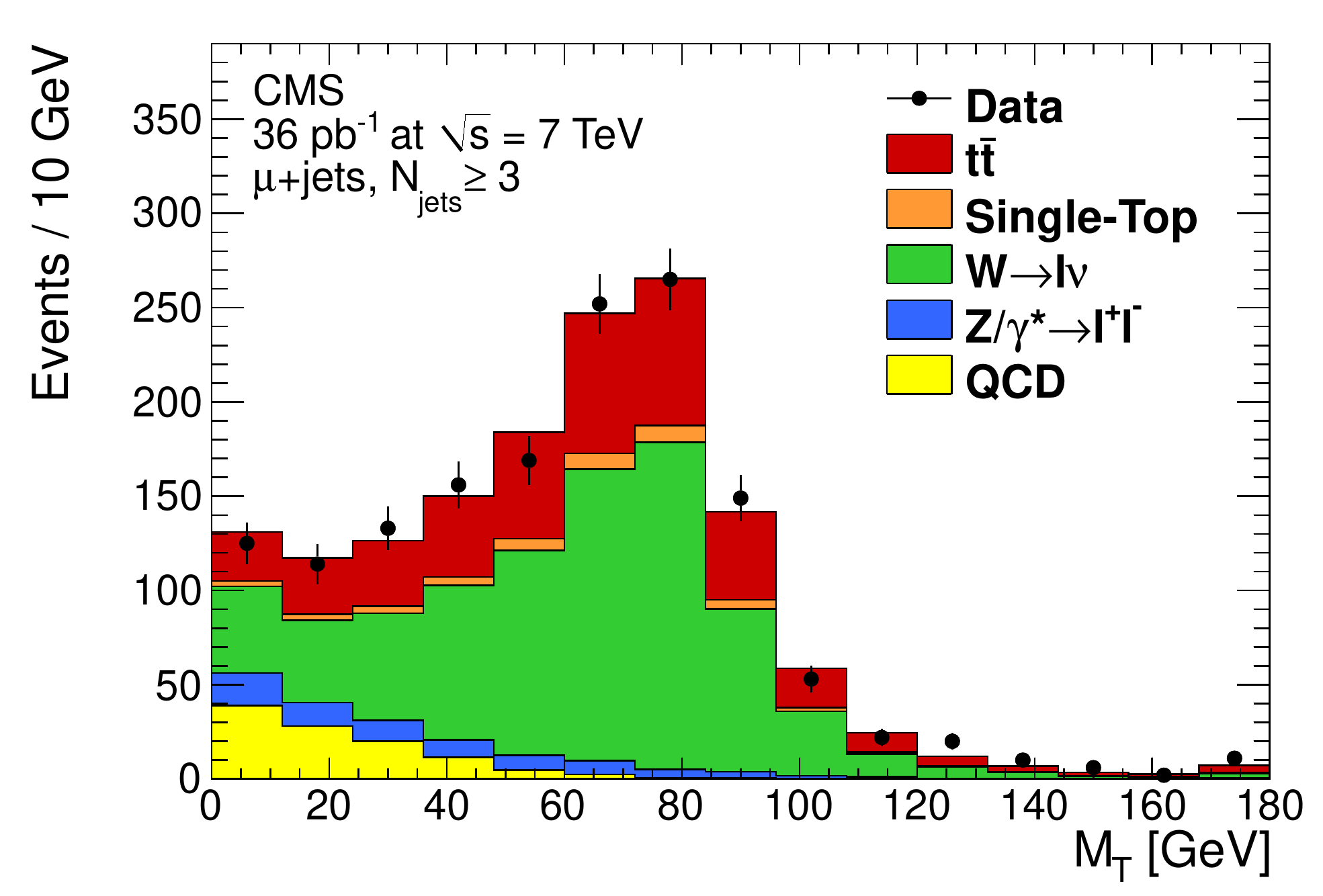}\\[5mm]
\caption{Transverse mass ($\rm{M}_{\mathrm{T}}$) distributions from data and simulation (scaled to the fit results) for (left) electron$+$jets and (right) muon$+$jets inclusive three-jet samples.
}
   \label{fig:transverse_mass}
\end{figure}

The combined transverse mass of the charged lepton and the \met~is a
kinematic variable that lacks the discriminating power of the M3 and
\met~variables for identifying ${\rm t{\bar t}}$ decays.  However,
this variable does provide separation between events containing a
decaying $W$~boson and non-$W$-boson decays, and thus serves as an
independent check of the kinematics of the simulated samples used in
this analysis.  Distributions of the transverse mass in the
muon$+$jets and electron$+$jets channels are shown in
Fig.~\ref{fig:transverse_mass}
for events with three or more jets. Good agreement is found between
the data and the sum of the signal and background derived from
the simulation scaled to the fit results.   The reduced $\chi^2$ value from
a fit of the data to the simulation is 1.8 (0.7) in the
electron$+$jets (muon$+$jets) channel.

\subsection{Cross-checks}

To test the robustness of the result,
the ${\rm t{\bar t}}$ cross section
is also determined in the muon$+$jets channel using four additional methods.
In the first method, 
we use a procedure based on counting the
number of events with an isolated muon and four or more jets.
This method uses an
event selection slightly different
from that described above.
Specifically, the jet ${p_\mathrm{T}}$ is required to be
greater than 25~$\mathrm{GeV}$ instead of 30~$\mathrm{GeV}$, and
the muon is required to have relative isolation less than
0.1, compared to 0.05 in the nominal selection.
Also the backgrounds from W/Z$+$jets
and QCD multijet events are calculated by using the technique of Berends
scaling~\cite{Berends}.
In the second method, we measure 
the ${\rm t{\bar t}}$ cross section using
a simultaneous fit to the distributions of jet multiplicity ($N_{\rm jets}$) and the
muon transverse momentum, ${p_{\rm{T}}}^\mu$. The jet multiplicity has
been shown (Table~\ref{tab:Eventyields}) to be a powerful variable for
separating top from QCD multijet and W$+$jets events.
The variable ${p_{\rm{T}}}^{\mu}$ is an attractive choice because it
is not directly affected by such systematics as
the JES and JER uncertainties. Furthermore, because the muon in either W or ${\rm t{\bar t}}$
production
comes from a W~decay, it receives a significant contribution to its
momentum from the W~rest mass, while muons from QCD multijet events
receive no such
boost.  In the third method, 
the ${\rm t{\bar t}}$ cross section
is determined
from a fit to the muon pseudorapidity distribution in order to separate the top
signal from the W$+$jets and QCD multijet backgrounds. This analysis uses
the asymmetry between inclusive W$^+$ boson and  W$^-$ boson
production, caused by the difference of the quark charges in the initial-state protons, to determine the templates for the
W$+$jets background.
A fourth method 
measures the
${\rm t{\bar t}}$ cross section from
events containing
a high-${p_{\rm T}}$ isolated muon and at least three jets.
For this analysis, we relax the relative isolation requirement to
$I_{\mathrm{rel}} < 0.1$ but introduce a requirement that the \met in the event is greater than 20~GeV, in order to keep the amount of
QCD multijet background small.
A method based on Ref.~\cite{D0MM,EvtClassWeight}
is used to estimate the amount of QCD multijet background separately for
events with three jets and events with at least four jets.
The number
of top-pair and W$+$jets events is extracted from a fit to the M3 distribution.
All four of these methods
give results consistent with our previously quoted measurement, but with
slightly larger combined statistical and systematic uncertainties in each case.

\begin{figure*}[h!]
  \begin{center}
    \includegraphics[width=0.49\textwidth]{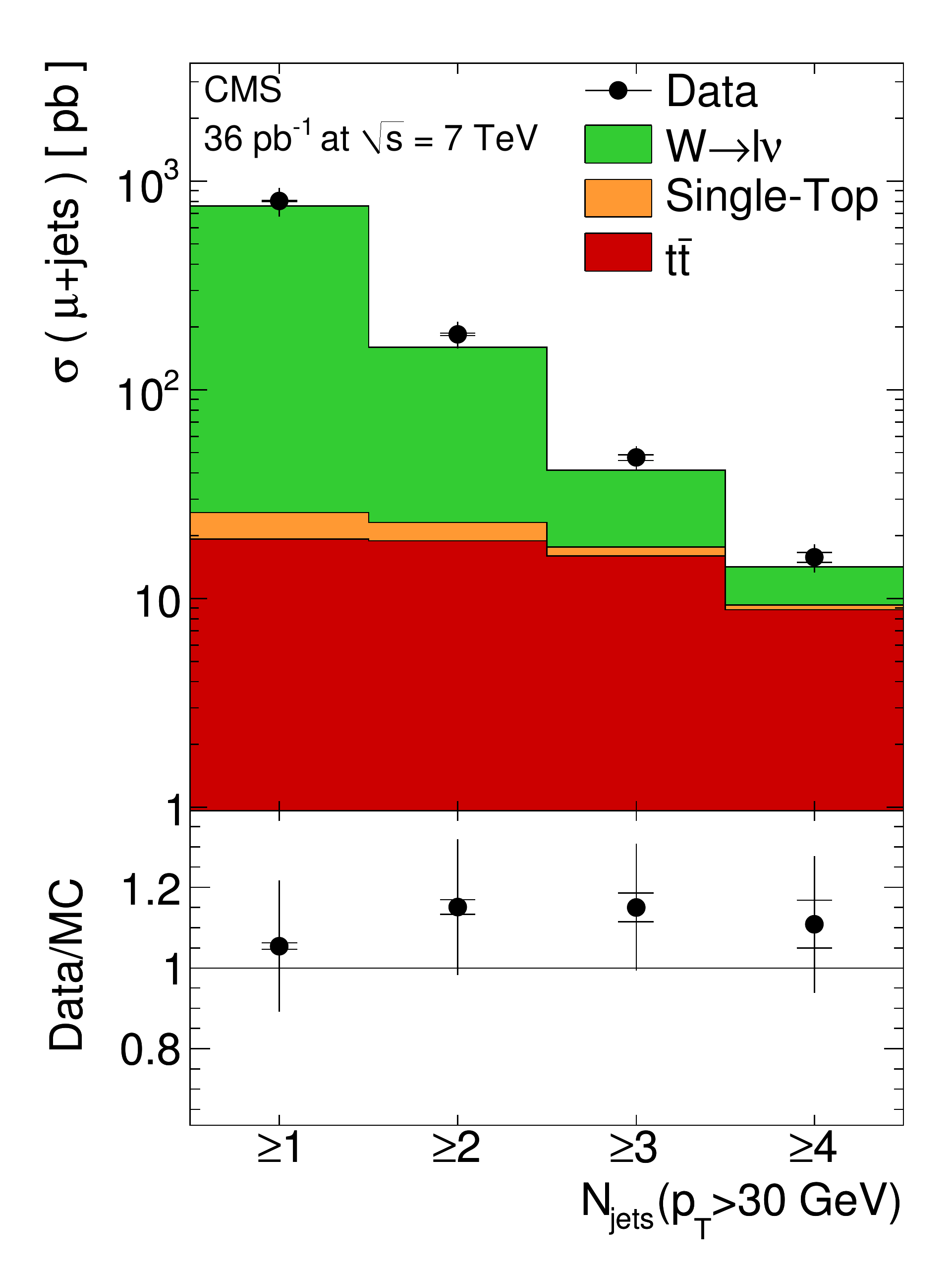}
    \includegraphics[width=0.49\textwidth]{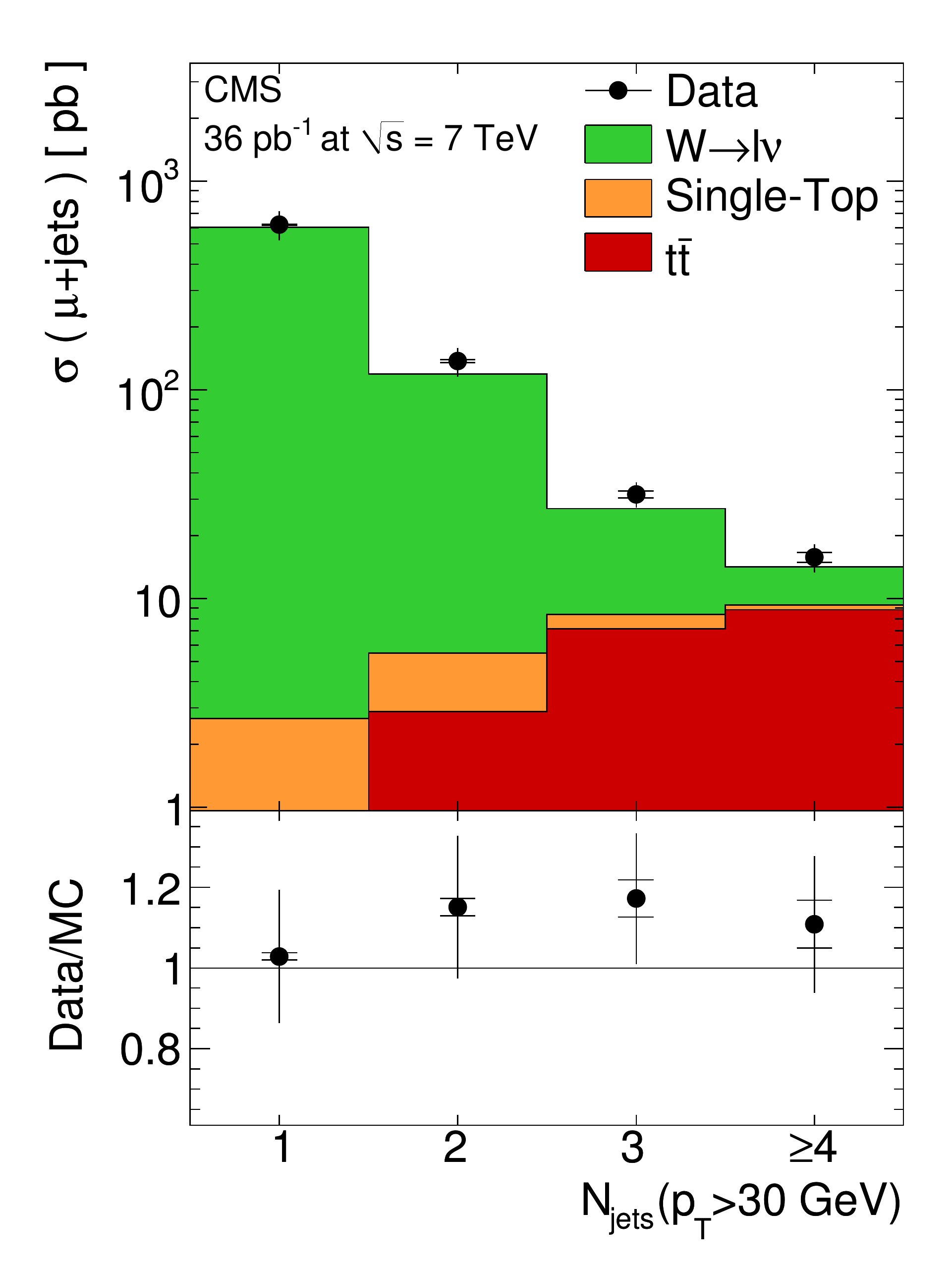}
    \setlength{\unitlength}{\textwidth}
    \begin{picture}(0,0)
         \end{picture}
    \caption {Cross section for the production of an isolated muon originating
from a W boson (including decays via tau
      leptons in the intermediate state) in association with additional
hard jets as a function of the
      (left) inclusive and the (right) exclusive multiplicity $N_{\rm jets}$
of jets with $p_{\rm T}>30$~GeV in the visible range
      of the detector. The inner error bars on the data points
correspond to the statistical uncertainties while the
      full error bars correspond to the statistical and systematic
uncertainties added in quadrature. The data points
      are compared with the expectation from the event generators used
for the simulation. The scaling factors derived in the
main analysis are not applied.}
    \label{fig:CrossSectionNJet}
  \end{center}
\end{figure*}

Complementary to the top-quark-pair production measurements, the
cross section for the production of exactly one
muon in association with additional hard jets is measured. In all
processes considered as signal for
this measurement, the muon originates from a W~boson. 
Both single top quark decays and
decays of top-quark pairs in the lepton$+$jets channel, including
decays via tau leptons in the intermediate state, are contributors to
this signature.  An additional component of this signal comes from the
production of a W boson with additional jets, which is the most prominent
background for the analysis of ${\rm t{\bar t}}$ ``lepton$+$jets'' decays.
The same event selection as described in Section~\ref{sec:muon_jets} is
applied. In addition, all jets in data are corrected for pileup, leading to reduced
JES and pileup uncertainties. To
obtain the cross section, the observed number of events in data is
corrected for the remaining background
processes. These include QCD multijet production, the production of a Z
boson with additional jets, single-top-quark decays, and other
${\rm t}\bar{\rm t}$ decays. The number of QCD multijet
events is determined from data using a template fit to the
missing-transverse-energy distribution in
each inclusive (or exclusive) jet-multiplicity bin. The normalization and
shape of the other backgrounds is
taken from the simulation.
Figure~\ref{fig:CrossSectionNJet} shows the cross section for the
production of a single muon with
\pt$>20~\mathrm{GeV}$ and $|\eta|<2.1$ and additional jets as a
function of the
inclusive and exclusive
multiplicity of jets with \pt$>30~\mathrm{GeV}$
within $|\eta|<2.4$.
The transition
from a phase space dominated by W$+$jets events (in the 1-jet and 2-jet
bins) towards a region dominated by
the production of top-quark pairs (in the 4-jet bin) is clearly visible.
The comparison of data and simulation
indicates a good understanding of this transition, while the overall
normalization seems to be slightly
underestimated. This is consistent with the main analysis, which also
found a W$+$jets cross section larger than the theoretical prediction.

\section{Discussion and Conclusion}
\label{sec:conclusion}

A measurement of the cross section
for top-quark pair production in proton-proton
collisions at a centre-of-mass energy of 7~\rm{TeV}
has been performed at the LHC
with the CMS detector.
The analysis uses
a data sample corresponding to an integrated luminosity
of 36~pb$^{-1}$ and
is based on the reconstruction of final states containing one
isolated, high transverse-momentum muon
or electron
and hadronic jets.
The measured cross section for the combined electron$+$jets
and muon$+$jets channels is
$173^{+39}_{-32}\,{\rm(stat. + syst.)} 
\pm 7\,{\rm (lumi.)}$~pb.
This measurement agrees with but has a
larger uncertainty than
current theoretical 
values~\cite{mcfm, mcfm:tt, Aliev, Langenfeld, Kidonakis},
which agree among themselves.
For example, 
the approximate NNLO calculation from Ref.~\cite{Kidonakis} yields
$163^{+7}_{-5}\,({\rm scale}) \pm 9\,({\rm PDF})\,{\rm pb}$, while a
similar calculation performed using HATHOR~\cite{Aliev, Langenfeld}
yields $160^{+5}_{-9}\,({\rm scale}) \pm 9\,({\rm PDF})\,{\rm pb}$.
For this calculation,
${Q^2}=\rm{(173\,GeV)^2}$ is chosen for both the factorization and
renormalization scales and the
MSTW2008 NNLO~\cite{mstw08} PDF set is used. 
The scale uncertainty is evaluated by
independently varying the scales by factors of 4 and 0.25, and the PDF
uncertainty is calculated using the 90\%\ confidence level envelope of
the PDF~\cite{Aliev, Langenfeld}.
Our cross section measurement also agrees with the earlier
CMS measurement in
the dilepton channel~\cite{CMS_dilepton} 
and the 
ATLAS measurement in the combined dilepton and
lepton$+$jets channels~\cite{ATLAS_top_sigma},
but has a smaller uncertainty than either of these previous results.
Given the agreement between theory and the 
experimental measurements in both the
dilepton and lepton+jets channels, no sign of new physics has emerged
in these studies, 
and the top quark at the LHC remains 
consistent with being a very massive particle whose
properties are as predicted by the standard model.

\section*{Acknowlegements}

\hyphenation{Bundes-ministerium Forschungs-gemeinschaft
  Forschungs-zentren} We wish to congratulate our colleagues in the
CERN accelerator departments for the excellent performance of the LHC
machine. We thank the technical and administrative staff at CERN and
other CMS institutes. The cost of the detectors, computing
  infrastructure, data acquisition and all other systems without which
  CMS would not be able to operate was supported by the financing
  agencies involved in the experiment. We are particularly indebted
  to: the Austrian Federal Ministry of Science and Research; the
  Belgium Fonds de la Recherche Scientifique, and Fonds voor
  Wetenschappelijk Onderzoek; the Brazilian Funding Agencies (CNPq,
  CAPES, FAPERJ, and FAPESP); the Bulgarian
Ministry of Education and Science; CERN; the Chinese Academy of
Sciences, Ministry of Science and Technology, and National Natural
Science Foundation of China; the Colombian Funding Agency
(COLCIENCIAS); the Croatian Ministry of Science, Education and Sport;
the Research Promotion Foundation, Cyprus; the Estonian Academy of
Sciences and NICPB; the Academy of Finland, Finnish Ministry of
Education, and Helsinki Institute of Physics; the Institut National de
Physique Nucl\'eaire et de Physique des Particules~/~CNRS, and
Commissariat \`a l'\'Energie Atomique et aux \'Energies
Alternatives~/~CEA, France; the Bundesministerium f\"ur Bildung und
Forschung, Deutsche Forschungsgemeinschaft, and Helmholtz-Gemeinschaft
Deutscher Forschungszentren, Germany; the General Secretariat for
Research and Technology, Greece; the National Scientific Research
Foundation, and National Office for Research and Technology, Hungary;
the Department of Atomic Energy, and Department of Science and
Technology, India; the Institute for Studies in Theoretical Physics
and Mathematics, Iran; the Science Foundation, Ireland; the Istituto
Nazionale di Fisica Nucleare, Italy; the Korean Ministry of Education,
Science and Technology and the World Class University program of NRF,
Korea; the Lithuanian Academy of Sciences; the Mexican Funding
Agencies (CINVESTAV, CONACYT, SEP, and UASLP-FAI); the Pakistan Atomic
Energy Commission; the State Commission for Scientific Research,
Poland; the Funda\c{c}\~ao para a Ci\^encia e a Tecnologia, Portugal;
JINR (Armenia, Belarus, Georgia, Ukraine, Uzbekistan); the Ministry of
Science and Technologies of the Russian Federation, and Russian
Ministry of Atomic Energy; the Ministry of Science and Technological
Development of Serbia; the Ministerio de Ciencia e Innovaci\'on, and
Programa Consolider-Ingenio 2010, Spain; the Swiss Funding Agencies
(ETH Board, ETH Zurich, PSI, SNF, UniZH, Canton Zurich, and SER); the
National Science Council, Taipei; the Scientific and Technical
Research Council of Turkey, and Turkish Atomic Energy Authority; the
Science and Technology Facilities Council, UK; the US Department of
Energy, and the US National Science Foundation.  

Individuals have received support from the Marie-Curie programme and
the European Research Council (European Union); the Leventis
Foundation; the A. P. Sloan Foundation; the Alexander von Humboldt
Foundation; the Associazione per lo Sviluppo Scientifico e Tecnologico
del Piemonte (Italy); the Belgian Federal Science Policy Office; the
Fonds pour la Formation \`a la Recherche dans l'Industrie et dans
l'Agriculture (FRIA-Belgium); and the Agentschap voor Innovatie door
Wetenschap en Technologie (IWT-Belgium).

\bibliography{auto_generated}
\cleardoublepage\appendix\section{The CMS Collaboration \label{app:collab}}\begin{sloppypar}\hyphenpenalty=5000\widowpenalty=500\clubpenalty=5000\textbf{Yerevan Physics Institute,  Yerevan,  Armenia}\\*[0pt]
S.~Chatrchyan, V.~Khachatryan, A.M.~Sirunyan, A.~Tumasyan
\vskip\cmsinstskip
\textbf{Institut f\"{u}r Hochenergiephysik der OeAW,  Wien,  Austria}\\*[0pt]
W.~Adam, T.~Bergauer, M.~Dragicevic, J.~Er\"{o}, C.~Fabjan, M.~Friedl, R.~Fr\"{u}hwirth, V.M.~Ghete, J.~Hammer\cmsAuthorMark{1}, S.~H\"{a}nsel, M.~Hoch, N.~H\"{o}rmann, J.~Hrubec, M.~Jeitler, W.~Kiesenhofer, M.~Krammer, D.~Liko, I.~Mikulec, M.~Pernicka, H.~Rohringer, R.~Sch\"{o}fbeck, J.~Strauss, A.~Taurok, F.~Teischinger, P.~Wagner, W.~Waltenberger, G.~Walzel, E.~Widl, C.-E.~Wulz
% ...
\vskip\cmsinstskip
% ...
1:~~Also at CERN, European Organization for Nuclear Research, Geneva, Switzerland\\
% ...

\end{sloppypar}
\end{document}